\documentclass[journal,final]{IEEEtran}
\usepackage{amsfonts}
\usepackage{amssymb}
\usepackage{amsmath,lipsum}
\usepackage{graphicx}
\usepackage{color}
\usepackage{multirow}
\usepackage{cite}
\usepackage{verbatim}
\usepackage{bm}
\usepackage{epstopdf}
\usepackage{algorithm, algorithmic}
\input{tcilatex}
\newtheorem{property}{Property}

\newtheorem{theo}{Theorem}

\begin{document}

\title{Lattice-Code Multiple Access: Architecture and Efficient Algorithms}
\author{ Tao Yang,~\IEEEmembership{Member,~IEEE}, Fangtao Yu, Rongke Liu,~%
\IEEEmembership{Senior Member,~IEEE}, Shanxiang Lyu,~%
\IEEEmembership{Member,~IEEE} and John Thompson,~%
\IEEEmembership{Fellow,~IEEE} \thanks{%
This work is supported by the National Natural Science Foundation (No.
62371020), Natural Science Foundation of Beijing (No. L232044) and National Key R\&D Program of China (No.2020YFB1807102 and No.
2022YFB2902604). This work was partly presented in IEEE Globecom 2023. } }
\maketitle

\begin{abstract}
This paper studies a $K$-user lattice-code based multiple-access (LCMA) scheme. Each user equipment (UE) encode its message with a practical lattice code, where we suggest a $2^m$-ary
\emph{ring code} with symbol-wise bijective mapping to $2^m$-PAM. The coded-modulated signal is spread with its designated signature
sequence, and all $K$ UEs transmit
simultaneously. The LCMA receiver choose some integer coefficients, computes the associated $K$ streams of \emph{integer
linear combinations} (ILCs) of the UEs' messages, and then reconstruct all UEs' messages from these ILC streams. To execute this, we put forth new efficient
LCMA \emph{soft detection} algorithms, which calculate the a posteriori probability
of the ILC over the lattice. The complexity is of order no
greater than $O(K)$, suitable for massive access of a large $K$.
The soft detection outputs are forwarded to $K$ ring-code decoders, which employ $2^m$-ary belief propagation to
recover the ILC streams. 

To identify the optimal integer coefficients of the ILCs, a new ``%
\emph{bounded independent vectors problem}" (BIVP) is established. We then solve this
BIVP by developing a new \emph{rate-constraint sphere decoding} algorithm,
significantly outperforming existing LLL and HKZ lattice reduction methods.
Then, we develop optimized signature sequences of LCMA using a new
target-switching steepest descent algorithm. With our developed algorithms,
LCMA is shown to support a significantly higher load of UEs and exhibits
dramatically improved error rate performance over state-of-the-art multiple
access schemes such as interleave-division multiple-access (IDMA) and
sparse-code multiple-access (SCMA). The advances are achieved with just
parallel processing and $K$ single-user decoding operations, avoiding the
implementation issues of successive interference cancelation and iterative
detection.
\end{abstract}


\markboth{}{Shell
\MakeLowercase{\textit{et al.}}: Bare Demo of IEEEtran.cls for IEEE Journals}


\section{Introduction}

The \textit{multiple access} (MA) problem is about how to support reliable
communication of $K$ user equipments (UEs)' within $N$ resource blocks in
time, frequency and spatial domains. For very large values of $K$, it
becomes a \textit{massive access} problem that is essential to
\textquotedblleft ubiquitous massive connectivity\textquotedblright\
envisaged for 6G \cite{6GTextbookTongWen}. It is well-known that orthogonal
MA is fundamentally limited from the following perspective. First, the
number of supported UEs $K$ is capped by the number of resource blocks $N$.
Second, dynamic resource allocation is required to maintain the
orthogonality, where the signaling cost could skyrocket as $K$ becomes
large. Third, despite the orthogonalization at the transmitters, the
wireless channel induces signal distortion that can easily destroy the
orthogonality \cite{Dai18,ding2017application}. Thus, orthogonal MA may not be a
suitable choice for massive access.

Non-orthogonal MA (NoMA) allows collisions of multiple UEs' packets. As
such, the number of UEs $K$ can go beyond the number of resource blocks $N$%
\cite{CoverTextbook}. Further, one can trade-in a higher $K$ by reducing the
peak rates of individual UEs, providing a high flexibility that are very
much desired for ubiquitous MA \cite{6GTextbookTongWen}. Moreover, NoMA
enables grant-free (GF) transmission, with which the signalling overhead
incurred by dynamic resource allocation can be slashed, making it possible
to realize massive access in 6G.
\subsection{Motivations}
The core issue of MA is how to deal with multi-UE interference (MUI). Most
existing NoMA schemes are based on \emph{interference suppression} and \emph{%
cancelation}, as in power-domain NoMA
with successive interference cancelation (SIC) and code-domain NoMA with
iterative detection and decoding (IDD) \cite{ten2003design}. In theory, such
mechanism generally leads to a reduced system load ($K/N)$.
In practice, SIC and IDD are subject to issues such as rate loss, error
propagation, slow and unguaranteed convergence, which have prevented them
from being implemented in 5G. This motivates us to study efficient MA approaches that can avoid the issues SIC and IDD in existing NoMA.

\subsection{Contributions}

In this paper, we suggest a lattice-code based MA (LCMA) system. In contrast to
conventional NoMA schemes that suppress and cancel MUI, LCMA embraces MUI by exploiting
the mapping between the structure of $K$ UEs' superimposed signal and the
lattice space. In the uplink, $K$ user equipments (UEs) encode their
messages with a $2^m$-ary \emph{ring code}, which are then bijectively mapped to $2^m$-PAM symbol-by-symbol. Such coded-modulation belongs to
the ensemble of \emph{lattice codes} with easy implementation. Each UE's coded-modulated signal is spread with its
designated signature sequence, specifically designed for LCMA, and all UEs
transmit simultaneously. The LCMA receiver choose some integer coefficients, computes the associated $K$ streams of \emph{integer
linear combinations} (ILCs) of the UEs' messages, and then reconstruct all UEs' messages from these ILC streams. This
paper contributes to this subject by developing a package of efficient algorithms
involving:

1) Simple yet powerful lattice codes that are in line with the mainstream $%
2^m$-QAM modulation.

2) Efficient LCMA soft detection algorithms whose per-UE complexity is less
than $O(K)$.

3) A new rate-constrained sphere-decoding algorithm that identifies the optimized LCMA integer coefficients.

4) A new target-switching steepest descent algorithm for the optimized LCMA signature sequences.

We demonstrate that the developed LCMA system exhibits much higher MA system
loads $K/N$ and lower error rates over baseline NoMA schemes such as
interleave-division MA (IDMA) and sparse-code MA (SCMA). For example, LCMA
achieves system loads of up to $K/N=350\%$ in Gaussian MA channel and
multi-user MIMO channel, which dramatically outperforms IDMA and SCMA that
can barely achieve $K/N=200\%$. Meanwhile, ultra-low block error rate (BLER)
of $10^{-6}$ to $10^{-7}$ is demonstrated for LCMA, which is rarely seen in
conventional MA schemes. Such advanced MA functionality and performance are
achieved with low-latency parallel processing, low detection complexity of
order less than $O(K)$ per-user, and exactly $K$ channel-code decoding
operations, without the need of SIC or IDD. Also, off-the-shelf channel
codes such as 5G NR LDPC codes can be directly used in LCMA for any system
load, avoiding the issue of adaptation of channel-code and multi-user
detector in IDMA.

\subsection{Related Literature}

\subsubsection{Multiple-access}

MA schemes based on interference cancelation and suppression have
been studied in the past two decades \cite%
{islam2016power,ding2017application}. Not long after the discovery of turbo
codes in 1993, the \textquotedblleft turbo principle" was introduced for the
multi-user decoding, first by Wang \& Poor \cite{wang1999iterative}. Since
2000, turbo-like IDD has been extensively researched. In \textquotedblleft
turbo-CDMA\textquotedblright\ \cite{wang1999iterative}, the inner code is a
multi-user detector with soft interference cancelation and linear minimum
MSE (MMSE) suppression, while the outer code is a bank of $K$ convolutional
code decoders. Soft probabilities are exchanged among these components
iteratively. In 2006, Li \textit{et al.} introduced a chip-level interleaved
CDMA, named interleave-division multiple-access (IDMA) \cite{LiTWC06}. The
chip interleaver enables uncorrelated chip interference, and thus a simple
matched filter optimally combines the chip-level signal to yield the
symbol-level soft information.

Low-density spreading CDMA and sparse-code MA (SCMA) differ from IDMA in
that each symbol-level signal is spread only to a small number of chips,
which forms a sparse matrix in the representation of the multi-user signal
that can be depicted using a bi-partite factor graph \cite%
{nikopour2013sparse}. SCMA also supports grant-free (GF) MA mode for the
massive-connectivity scenario. Spatially coupled codes were also studied for
dealing with the MA problem, yielding enlarged admissible region for fading
MA channels \cite{KudekarISIT11}. For IDMA and SCMA, spreading/sparse codes
with irregular degree profiles were investigated including the work of
ourselves \cite{YangTWC09,nikopour2013sparse}, which yielded improved
convergence behavior of the multi-user decoding.

Rate-splitting MA (RSMA) was studied for closed-loop systems \cite%
{RimoldiIT96,MaoTcom19}. The idea is to superimpose a common message on the
private messages, which may enlarge the rate-region. Other code-domain NoMA
techniques are proposed such as pattern division MA (PDMA), multi-user
shared access (MUSA) etc. \cite{chen2016pattern}, which exhibits some
advantages for implementation. For grant-free MA, active user identification
based on compressive sensing and coded slotted Aloha protocols are studied
\cite{ChengJSAC21,SunTcom17,PaoliniIT15}, which significantly reduces the
signaling overhead that is essential to massive access. Here we are not able
to list all existing results in the area of MA, and readers are encouraged
to refer to the excellent survey in \cite{Dai18}. Note that most existing MA
schemes rely on the notion of \textquotedblleft rejecting MUI", where the
MUI structure is not or insufficiently exploited. 

\subsubsection{Literature of Lattice-codes}

For general multi-user networks, it has been proved that \textquotedblleft {%
structured codes}" based on lattices can achieve a larger capacity region
compared to conventional ``random-like coding''. The proof
was based on the idea of \textquotedblleft algebraic binning" of codewords,
where each bin collects a certain subset of all codewords. The structure of
lattice-codes enables efficient generation of the bin-indices as in the
source coding with side information (SI) problem, and efficient decoding of
the bin-indices as in the channel coding with SI problem\cite{ZamirTIT02}.
For physical-layer network coding (PNC) or compute-forward (CF), by
adopting lattice-codes at source nodes, the receiver can directly compute
the bin-indices in the form of \emph{integer-combinations} of all users'
messages\cite{NazerIT11}, leading to significant coding gain or even
multiplexing gain\cite{NamIT10,YuanYangIT13}. The work \cite{LimTIT20}
studied simultaneous
computation of more than one integer-combinations. The results on using
lattice-codes for tackling MIMO detection and downlink MIMO precoding
problems were reported in \cite{ZhanIT14} and \cite{SilvaTWC17} under the
name of integer-forcing (IF). The latter borrowed the notion of reverse CF
which exploited the uplink-downlink duality \cite{HongIT13,YangTWC17}.
Various lattice reductions methods for identifying a \textquotedblleft good"
coefficient matrix for the integer-combinations have been reported in many
works such as \cite{SakzadTWC13}. Recently, CF and IF have been extended to
time-varying or frequency-selective fading channels using multi-mode IF and
ring CF \cite{YangTWC20,LyuTIT19}. The IF notion was also applied to solve
the inter-symbol-interference equalization problem with the help of cyclic
linear codes \cite{OrdentlichTIT12}. Here we are not able to list all
existing results on lattice-codes, CF and IF, and highly motivated readers
are encouraged to refer to \cite{ZamirTIT02} and \cite{LimTIT20}.

\subsubsection{Lattice-codes and MA}

From an information theoretic perspective, Zhu and Gastpar showed that any
rate-tuple of the entire Gaussian MA capacity region can be achieved using a
lattice-code based approach, and the scheme was named compute-forward MA
(CFMA) \cite{zhu2016gaussian}. In contrast to random-like coding approaches
exploited in existing NoMA schemes, lattice-code based MA exhibits a greater
capacity region, which is achieved with low-cost single-user decoding. The
design of CFMA for the Gaussian MA channel with binary codes was studied in
\cite{sula2018compute}. Recently, we extend the result of \cite%
{zhu2016gaussian} and \cite{sula2018compute} to fading MA channel with
practical $q$-ary codes\cite{ChenISIT22,ChenTWC22}. To date, most of the related works on lattice-codes for MA have been
focusing on achievable rates by proving the existence of \textquotedblleft
good" nested lattice-codes, whereas the practical aspects are not yet
sufficiently researched. The methods developed in \cite{sula2018compute} and
\cite{ChenTWC22} do not apply to practical $2^{2m}$-QAM signaling and MIMO.
The impacts of lattice-codes on the key performance indicators such as the
system load, BLER, latency, complexity and etc., are still to be
investigated. In addition, for a large $K$, there lacks efficient algorithms
for both the soft detection and the identification of the coefficient matrix
with realistic implementation costs. This motivates us to develop a package
of practical coding, efficient signal processing algorithms and optimization
methods for lattice-code MA in this paper.
\section{System Model}

Consider an uplink MA system where $K$ single-antenna UEs deliver messages
to a common base station (BS). Let a row vector $\mathbf{x}_{i}^{T}$ denote
the transmitted coded symbol sequence of UE $i$, $i=1,...,K$, with a
normalized average energy per-symbol.

\subsection{Scenario I. Gaussian Multiple Access}

For a Gaussian MA channel (G-MAC), the received signal at the single-antenna
BS is given by a row vector
\begin{equation}
\mathbf{y}^{T}=\sum_{i=1}^{K}\sqrt{\rho _{i}}\mathbf{x}_{i}^{T}+\mathbf{z}%
^{T}
\end{equation}%
where $\mathbf{z}^{T}$ denotes the additive white Gaussian noise (AWGN)
sequence whose entries are i.i.d. with mean 0 and variance $\sigma
_{z}^{2}=1 $. The average per-UE signal-to-noise ratio (SNR) is given by $%
\rho $, where $\rho =\frac{1}{K}\dsum\limits_{i=1}^{K}\rho _{i}$. The G-MAC model has high interests from a theoretical point of view. Also,
such model applies to line-of-sight dominant practical systems, e.g. a
satellite communication where a single-beam is allocated for a certain
geographical area containing a large number of UEs\footnote{In this paper we assume
that the signals of the UEs are synchronized at the receiver. This is done
in the initial access stage prior to the data transmission stage considered
in this paper.}. 


\subsection{Scenario II. Multi-UE (MU) MIMO}

For a (narrow-band) MU-MIMO channel, the baseband equivalent discrete signal
received at the $N_{R}$-antenna BS can be presented by the following real-valued
model
\begin{equation}
\mathbf{Y}=\sum_{i=1}^{K}\mathbf{h}_{i}\sqrt{\rho }\mathbf{x}_{i}^{T}+%
\mathbf{Z=}\sqrt{\rho }\mathbf{HX}+\mathbf{Z}  \label{Eq_SystemModel_MUMIMO}
\end{equation}%
where the column vector $\mathbf{h}_{i}$ denotes the channel coefficients
from UE $i$ to the $N_{R}$ antennas of the BS; the $j$th row of $\mathbf{Y}$
denotes the signal sequence received by antenna $j$, $j=1,...,N_{R}$; $%
\mathbf{Z}$ denotes the AWGN matrix. For the ease of presentation, identical
power among UEs is utilized in (\ref{Eq_SystemModel_MUMIMO}), while our
treatment is also applicable to unequal power allocation.

A complex-valued model can be represented by a real-valued model of doubled
dimension of $2N_{R}$-by-$2K$, i.e.,
\begin{equation}
\left[
\begin{array}{c}
\mathbf{Y}^{\func{Re}} \\
\mathbf{Y}^{\func{Im}}%
\end{array}%
\right] =\sqrt{\rho }\left[
\begin{array}{cc}
\mathbf{H}^{\func{Re}} & -\mathbf{H}^{\func{Im}} \\
\mathbf{H}^{\func{Im}} & \mathbf{H}^{\func{Re}}%
\end{array}%
\right] \left[
\begin{array}{c}
\mathbf{X}^{\func{Re}} \\
\mathbf{X}^{\func{Im}}%
\end{array}%
\right] \mathbf{+}\left[
\begin{array}{c}
\mathbf{Z}^{\func{Re}} \\
\mathbf{Z}^{\func{Im}}%
\end{array}%
\right]
\end{equation}%
following \cite{NazerIT11,YangTWC17_NoMA}. This paper presents with
real-valued model for the clarity of notation and better readability.

This paper is primarily interested in studying the configuration of $K\geq
N_R$. In particular, for $K$ being very large, the scenario is referred to
\textquotedblleft \textit{massive access MIMO}\textquotedblright. Such
configuration is different from conventional \emph{massive MIMO} in the
literature, where $N_R$ is very large while the number of UE is far less,
i.e. $K<<N_R$.  {Note that a fading MAC model, where each user's signal undergoes a channel gain and phase rotation, is equivalent to the MU-MIMO setup with $N_R=1$. Such fading MAC model is sitting in between the Gaussian MAC and MU-MIMO models described above.}

\subsection{Performance Indicators}
Following the convention in studying the uplink MA, we consider an open-loop
system where there is no feedback to the UE transmitters to deliver the
channel state information (CSI) and implement adaptive coding and modulation
(ACM). Each UE transmits at a target \emph{per-user data rate} $%
R_{0}$. The key performance indicators of the MA system are the \emph{%
supported system load} given by $K/N$, and \emph{block error rate} (BLER).

The problem under consideration is: how to
design a MA transceiver architecture and processing algorithms, such that
the MA system supports a high system load while meeting a target BLER, or
achieves a low BLER for a given system load.

\section{Architecture of Lattice-code Multiple access}

\begin{figure*}[tp]
\centering%
\includegraphics[width=6.4in,height=2.6in]{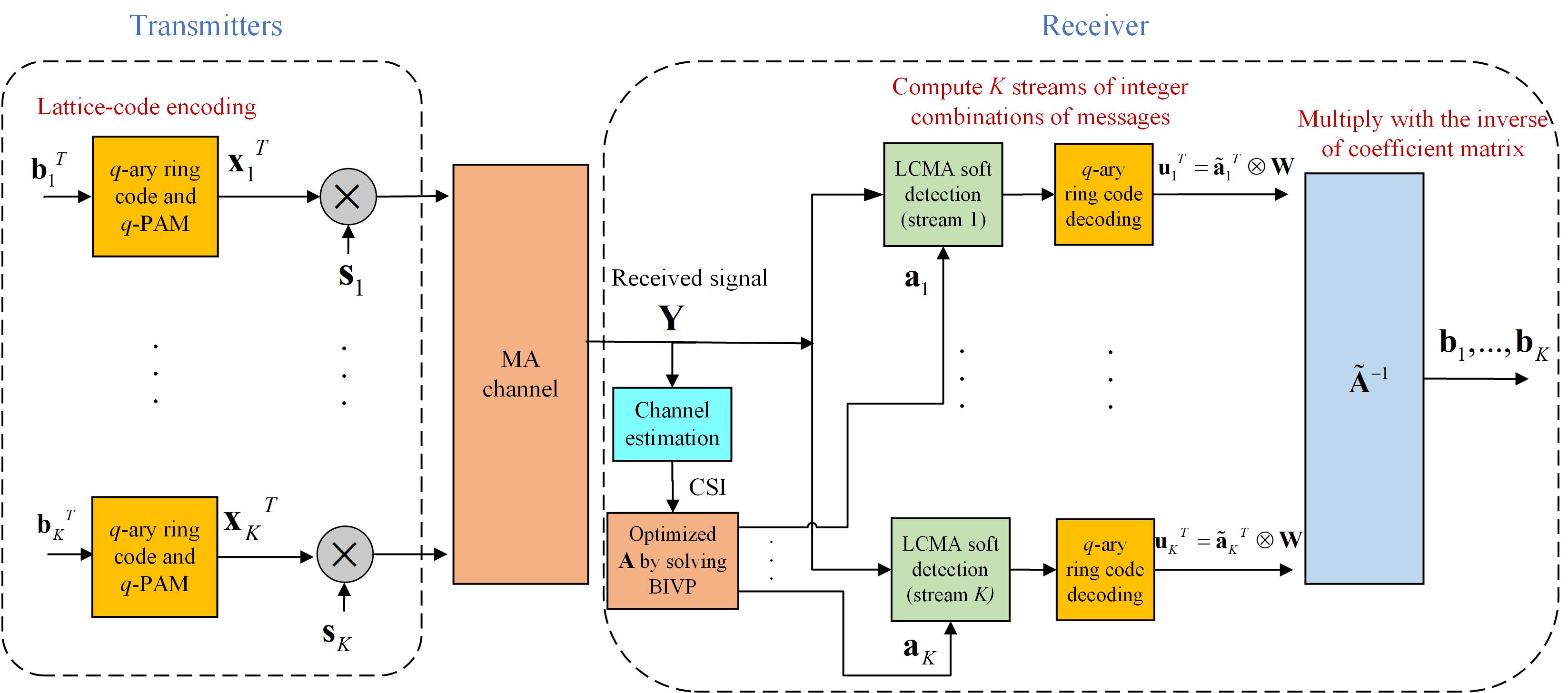}
\caption{Block diagram of the transmitters and receiver of a LCMA system.
All users utilize the same $2^{m}$-ary RCM. No interleavers and
deinterleavers are used. For each MA channel realization, the optimized
coefficient matrix $\mathbf{A}$ is identified by solving the BIVP w.r.t. the
channel state information. The LCMA soft detection and decoding of the $K$
streams are implemented in parallel. }
\label{Fig_LCMA_Tx}
\end{figure*}

The architecture of a LCMA system is depicted in Fig. \ref{Fig_LCMA_Tx}.
Our treatment applies to both G-MAC and MU-MIMO (as will be specified in Section IV. F).

\subsection{LCMA Transmitters}

\subsubsection{Encoding and Modulation with a Practical Lattice-code}

\

Let $\mathbf{b}_{i}\mathbf{=}\left[ b_{i}\left[ 1\right] ,\cdots ,b_{i}\left[
k\right] \right] ^{T}$ denote the message sequence of UE $i$, $i=1,...,K$.
Each entry of $\mathbf{b}_{i}$ belongs to an integer ring\footnote{%
The conversion from a binary message sequence to a $2^{m}$-ary message
sequence is straightforward. Our development applies to a $q$-ary code with $%
q$-PAM for any $q$, either prime or non-prime, while this paper only
presents with non-prime $q=2^{m}$.} $\mathbb{Z}_{2^{m}}\triangleq \left\{
0,\cdots ,2^{m}-1\right\} $, $m=1,2,\cdots $. For a general value of $m$, we
suggest to utilize a $2^{m}$-ary ring code, with \emph{generator matrix} $%
\mathbf{G}$ of size $n$-by-$k$, to encode the message sequences of the $K$
UEs. The resultant coded sequences are given by%
\begin{equation}
\mathbf{c}_{i}=\func{mod}\left( \mathbf{Gb}_{i}\mathbf{,}2^{m}\right) =%
\mathbf{G\otimes b}_{i},i=1,\cdots ,K,  \label{Eq_encodinggeneral}
\end{equation}%
where \textquotedblleft $\mathbf{\otimes }$\textquotedblright\ represents
matrix multiplication modulo-$2^{m}$.

The entries of each UE's coded sequence $\mathbf{c}_{i}=\left[ c_{i}\left[ 1%
\right] ,\cdots ,c_{i}\left[ n\right] \right] ^{T}$ are mapped \textit{%
one-to-one}\ to symbols in a $2^{m}$-PAM constellation, given by
\begin{equation}
x_{i}\left[ t\right] =\frac{1}{\gamma }\left( c_{i}\left[ t\right] -\frac{%
2^{m}-1}{2}\right) \in \frac{1}{\gamma }\left\{ \frac{1-2^{m}}{2},\cdots ,%
\frac{2^{m}-1}{2}\right\}.  \label{Eq_modulation}
\end{equation}%
Here $\gamma $ normalizes the average symbol energy.

The above 2$^{m}$-ary ring-coded PAM executed via (\ref{Eq_encodinggeneral})
and (\ref{Eq_modulation}) is a lattice code, which utilizes a simple
one-dimension shaping lattice\cite{ErezTIT05}\cite{ZamirTIT02}\cite{NamIT10}%
. Such a lattice code matches with the mainstream $2^{2m}$-QAM signaling%
\footnote{%
For a complex-valued model, two independent $2^{m}$-level ring-coded PAM,
one for the in-phase and the other for the quadrature part, form a lattice
code with $2^{2m}$-QAM signaling. For a better readability, this paper
presents with the real-valued model.}. 
The per-user rate of such a lattice code is $R_{0}=\frac{k}{n}\log _{2}2^{m}=%
\frac{km}{n}$ bits/symbol\footnote{%
The extension to the asymmetric rate setup is straightforward. A low rate
UE's message is zero-padded to form a length $k$ message sequence. Then, the
same channel code encoder is utilized to encode all UEs' messages.}. For $%
m=1 $, any existing binary code (such as LDPC or polar codes in the 5G
standard) applies, where $\mathbf{\ c}_{i}$ is a binary sequence while $%
\mathbf{x}_{i}=[x_{i}[1],\cdots ,x_{i}[n]]$ is a BPSK symbol sequence. For a
general $m $, LPDC codes and capacity approaching doubly irregular
accumulate codes over $2^{m}$-ary rings that we developed previously can be
utilized \cite{YuTcom22}.

\subsubsection{Spreading for MA}

We suggest to multiply the \emph{symbol-level} signal $x_{i}\left[ t\right] $
of UE $i$ with its designated spreading signature sequence $\mathbf{s}_{i}$
of length $N_{S}$, yielding the \emph{chip-level} signal $\mathbf{s}_{i}x_{i}%
\left[ t\right] $. The chip-level signal sequence can be written as%
\begin{equation}
\left[ \mathbf{s}_{i}^{T}x_{i}\left[ 1\right] ,\cdots,\mathbf{s}_{i}^{T}x_{i}%
\left[ n\right] \right] ,i=1,\cdots,K.
\end{equation}%
Then all $K$ UEs transmit simultaneously.

The spreading sequences satisfy a total power constraint%
\begin{equation}
\sum_{i=1}^{K}\left\Vert \mathbf{s}_{i}\right\Vert ^{2}\leq K.
\end{equation}%

Compared to other existing MA schemes, the distinguishing features of LCMA
transmitter involve: a lattice code formed via $2^{m}$-ary ring code coupled
with a one-to-one $2^{m}$-PAM mapping\footnote{%
This differs from conventional bit-interleaved coded modulation, trellis
coded modulation and multi-level coding schemes.}, a spreading matrix $%
\mathbf{S}=\left[ \mathbf{s}_{1},\cdots,\mathbf{s}_{K}\right] $ whose design
is coupled with the LCMA receiver's processing (to be presented next), and
the removal of the symbol-level or chip-level interleavers\cite%
{wang1999iterative}\cite{ping2006interleave}. Note that even for $m=1$,
lattice-based processing and design are required.

\subsection{LCMA Receiver}

The received signal at the BS is given by\footnote{For clarity, this paper presents with receiver-side synchronization setup. The developed LCMA also applies to the setup with asynchrony among the UEs.}
\begin{equation}
\mathbf{Y=}\sum_{i=1}^{K}\mathbf{s}_{i}^{T}\mathbf{x}_{i}+\mathbf{Z.}
\end{equation}%
The $j$th row stands for the signal w.r.t. the $j$th chip, $j=1,\cdots,N_{S}$%
. The task of the BS receiver is to recover all $K$ UE's message sequences $%
\mathbf{b}_{1},\cdots,\mathbf{b}_{K}$.

The receiver architecture is shown in Fig. \ref{Fig_LCMA_Tx}. The notion of
the LCMA's receiver processing is to efficiently compute $K$ independent
streams of \emph{integer linear combinations} (ILCs) of the UE' messages,
with the help of the structural property of the underlying lattice code.

\definition The $l$th stream of \emph{message-level} ILC is defined as
\begin{equation}
\mathbf{u}_{l}^{T}\mathbf{\triangleq }\func{mod}\left( \sum_{i=1}^{K}a_{l,i}%
\mathbf{b}_{i}^{T},2^{m}\right) =\mathbf{a}_{l}^{T}\mathbf{\otimes B}%
,l=1,\cdots ,K,
\end{equation}%
where $\mathbf{B}$ $=\left[ \mathbf{b}_{1},\cdots ,\mathbf{b}_{K}\right]
^{T} $, $\mathbf{a}_{l}=$ $\left[ a_{l,1},\cdots ,a_{l,K}\right] ^{T}$
denotes the associated integer \emph{coefficient vector}.

Let $\mathbf{U}=\left[ \mathbf{u}_{1},\cdots ,\mathbf{u}_{K}\right] ^{T}$
consist of all $K$ streams of ILC. Let $\mathbf{A=}\left[ \mathbf{a}%
_{1},\cdots ,\mathbf{a}_{K}\right] ^{T}$ stack up all $K$ coefficient
vectors, referred to as the ILC \textit{coefficient matrix}. Denote $\mathbf{%
A}$ modulo-$2^{m}$ by $\widetilde{\mathbf{A}}=\func{mod}\left( \mathbf{A,}%
2^{m}\right) $, then
\begin{equation}
\mathbf{U=\widetilde{\mathbf{A}}\otimes B.}
\end{equation}%
In LCMA, it is required that $\widetilde{\mathbf{A}}$ is of full rank $K$ in
$\mathbb{Z}_{2^{m}}$, thus it has a unique inverse $\widetilde{\mathbf{A}}%
^{-1}$, i.e., $\widetilde{\mathbf{A}}^{-1}\mathbf{\otimes }\widetilde{%
\mathbf{A}}=\mathbf{I}$. If all streams of ILCs $\mathbf{u}_{1},\cdots ,%
\mathbf{u}_{K}$ are correctly computed, all $K$ users' messages can be
recovered by implementing%
\begin{equation}
\mathbf{B=}\widetilde{\mathbf{A}}^{-1}\mathbf{\otimes U.}
\label{Eq_RecoverMessages}
\end{equation}

\remark Conventionally, the BS receiver performs UE-by-UE detection and
decoding w.r.t. $\mathbf{b}_{1},\cdots ,\mathbf{b}_{K}$. The LCMA receiver
goes beyond this. It first performs ILC-by-ILC detection and decoding w.r.t.
$\mathbf{u}_{1},\cdots ,\mathbf{u}_{K}$, and then retrieves the message
sequences via (\ref{Eq_RecoverMessages}). LCMA has the room to identify the
best matrix $\mathbf{A}$ that leads to the most efficient ILC-by-ILC
processing, which translates into enhanced system load or error rate
performance. When $\mathbf{A}$ is set to $\mathbf{I}$, LCMA receiver reduces
to conventional UE-by-UE processing.  {Note that the computation of the $K$ integer combinations are performed in parallel, where only the $l$-th row of $\mathbf{A}$ is needed in computing the $l$-th integer linear combination.}

\subsection{LCMA Design Problems to be Addressed}

For the LCMA receiver, the design problems involve:

$\bullet $ Efficient computation of $K$ ILC streams $\mathbf{U}=\left[
\mathbf{u}_{1},\cdots ,\mathbf{u}_{K}\right] ^{T}$ by exploiting the
structural property of the underlying lattice code. Practical algorithms for
this will be developed in Section IV.

$\bullet $ Identification of the optimal ILC coefficient matrix $\mathbf{A}$%
. This will be studied in Section V.A.

For the LCMA transmitters, the design problem is:

$\bullet $ Optimized design of the MA signature matrix $\mathbf{S}$. This
problem is coupled with the receiver processing, and will be studied in
Section V.B.

\section{Algorithms of Soft LCMA Detection and Decoding}

This section is devoted to the efficient computation of the ILC streams $%
\mathbf{U}=\left[ \mathbf{u}_{1},\cdots ,\mathbf{u}_{K}\right] ^{T}$, where $%
\mathbf{A}$ is given. We focus on a \emph{parallel rule\footnote{%
The parallel rule can be enhanced by a successive rule \cite{ChenTWC22}.
This paper focus on the parallel rule owing to its low-cost implementation,
tractable analysis and optimization. Later we will see that the parallel
rule yields competitive performance.}} for computing the a posteriori
probabilities (APPs) of the $K$ ILC streams, given by
\begin{equation}
p\left( \mathbf{u}_{l}|\mathbf{Y}\right) ,l=1,\cdots ,K.
\label{Eq_APP_parellel}
\end{equation}%
The execution of (\ref{Eq_APP_parellel}) relies on a crucial property of the
underlying lattice code presented below.

\subsection{Property of Lattice code Exploited in LCMA}

Recall the coded sequences $\mathbf{c}_{i}$ generated in (\ref%
{Eq_encodinggeneral}). Let $\mathbf{C=}\left[ \mathbf{c}_{1},\cdots ,\mathbf{%
c}_{K}\right] ^{T}$.

\definition The $l$th stream of \emph{codeword-level} ILC is defined as
\begin{equation}
\mathbf{v}_{l}^{T}\triangleq \func{mod}\left( \sum_{i=1}^{K}a_{l,i}\mathbf{c}%
_{i}^{T},2^{m}\right) =\mathbf{a}_{l}^{T}\mathbf{\otimes C},l=1,\cdots,K.
\end{equation}

\begin{property}
With the $2^{m}$-ary ring code utilized in (\ref{Eq_encodinggeneral}), we
have
\begin{equation}
\mathbf{v}_{l}
=\mathbf{G\otimes }\func{mod}\left( \sum_{i=1}^{K}a_{l,i}%
\mathbf{b}_{i},2^{m}\right) =\mathbf{G\otimes u}_{l}.  \label{Eq_Property1}
\end{equation}
\end{property}

For point-to-point communication, the codeword $\mathbf{c}$ and the message
sequence $\mathbf{b}$ are related by the generator matrix $\mathbf{G}$ as in
(\ref{Eq_encodinggeneral}). For the $K$-user MA setting under consideration,
Property 1 indicates that the codeword-level ILC $\mathbf{v}_{l}$ and the
message-level ILC $\mathbf{u}_{l}$ are also related by the generator matrix $%
\mathbf{G}$ as in (\ref{Eq_Property1}). (This property does not apply to
conventional non-lattice code based schemes such as bit-interleaved coded
modulation (BICM), trellis coded modulation (TCM) and superposition coded
modulation (SCM).)

Thanks to Property 1, a \textquotedblleft two-step\textquotedblright\ method
applies:

Step 1) \emph{\textbf{Soft ILC Detection}}: computes $p\left( \mathbf{v}_{l}|%
\mathbf{Y}\right) $, i.e., the APPs of the codeword-level ILC $\mathbf{v}%
_{l} $.

Step 2) \emph{\textbf{Ring-code Decoding}}: takes in $p\left( \mathbf{v}_{l}|%
\mathbf{Y}\right) $ as input for decoding, which outputs a decision on the
message-level ILC $\mathbf{u}_{l}$.


\subsection{LCMA Soft ILC Detection}

Here we consider Step 1) for the $l$-th ILC stream. The soft ILC detection
calculates
\begin{equation}
p\left( v_{l}\left[ t\right] |\mathbf{y}\left[ t\right] \right) ,t=1,\cdots,n
\label{Eq_APP_general}
\end{equation}%
in a symbol-wise manner. %
This paper focuses on a linear LCMA soft detector to calculate (\ref%
{Eq_APP_general}). (The non-linear LCMA detector is studied in a separate
work \cite{YangTaoCL23}.) The notion is to first transform the $N_{S}$%
-dimension received signal into $K$ streams of single-dimension signals.
Then, each stream is used to compute one ILC.

Let a length-$N_{S}$ vector $\mathbf{w}_{l}$ denote the linear filter w.r.t
the $l$th ILC stream, which is normalized to $\left\Vert \mathbf{w}%
_{l}\right\Vert =1$. For symbol-by-symbol detection, we can omit the symbol
index \textquotedblleft $t$\textquotedblright\ below. The $l$th filtered
signal stream is
\begin{eqnarray}
\widetilde{y}_{l} &=&\mathbf{w}_{l}^{T}\mathbf{y=w}_{l}^{T}\dsum%
\limits_{i=1}^{K}\sqrt{\rho }\mathbf{s}_{i}x_{i}+\widetilde{z}_{l} \\
&=&\dsum\limits_{i=1}^{K}\sqrt{\rho }\psi _{l,i}x_{i}+\widetilde{z}_{l}.
\end{eqnarray}%
where $\psi _{l,i}=\mathbf{w}_{l}^{T}\mathbf{s}_{i}$ denotes the
\textquotedblleft effective gain\textquotedblright\ w.r.t. UE $i$'s signal,
and the noise term $\widetilde{z}_{l}$ has a unit variance. Our developed
soft detection algorithm below applies to any $\mathbf{w}_{l}$.

Let the set $\mathcal{I}_{l}\triangleq \left\{ i:a_{l,i}\neq 0\right\} $
collects the positions of non-zero entries of $\mathbf{a}_{l}$, and $%
\mathcal{I}_{l}^{c}$ be the complementary set. Let $\omega \left( \mathbf{a}%
_{l}\right) \triangleq \left\vert \mathcal{I}_{l}\right\vert $ denote the
number of non-zero entries. Then, $\widetilde{y}_{l}$ is re-arranged as
\begin{equation}
\widetilde{y}_{l}=\dsum\limits_{i\in \mathcal{I}_{l}}\sqrt{\rho }\psi
_{l,i}x_{i}+\dsum\limits_{i\in \mathcal{I}_{l}^{c}}\sqrt{\rho }\psi
_{l,i}x_{i}+\widetilde{z}_{l}=\dsum\limits_{i\in \mathcal{I}_{l}}\sqrt{\rho }%
\psi _{l,i}x_{i}+\xi _{l}.
\end{equation}%
The term $\dsum\limits_{i\in \mathcal{I}_{l}}\sqrt{\rho }\psi _{l,i}x_{i}$
is the superposition of the signals of the $\omega \left( \mathbf{a}%
_{l}\right) $ users whose ILC coefficients are non-zero, which is the \emph{%
useful signal }part. The term $\dsum\limits_{i\in \mathcal{I}_{l}^{c}}\sqrt{%
\rho }\psi _{l,i}x_{i}$ contains the signals of the remaining $K-\omega
\left( \mathbf{a}_{l}\right) $ UEs whose ILC coefficients are zero, which
can be regarded as irrelevant w.r.t. ILC. The term $\xi _{l}=$ $%
\dsum\limits_{i\in \mathcal{I}_{l}^{c}}\sqrt{\rho }\psi _{l,i}x_{i}+%
\widetilde{z}_{l}$ is treated as the \emph{effective noise}, which is not
correlated with the useful signal part.

Recall the one-to-one mapping $x_{i}=\frac{1}{\gamma }\left( c_{i}-\frac{%
2^{m}-1}{2}\right) $ in (\ref{Eq_modulation}). For the clarity of
presentation, we express the received signal with $c_{i}$ (instead of with $%
x_{i}$), given as
\begin{equation}
\overline{y}_{l}
=\dsum\limits_{i\in \mathcal{I}_{l}}\sqrt{\rho }\psi _{l,i}\left(
\gamma x_{i}+\frac{2^{m}-1}{2}\right) +\gamma \xi _{l}=\dsum\limits_{i\in
\mathcal{I}_{l}}\sqrt{\rho }\psi _{l,i}c_{i}+\overline{z}_{l}.
\end{equation}

For a large $K$, it can be shown that $\left\vert \mathcal{I}%
_{l}^{c}\right\vert $ is sufficiently large to apply Central Limit Theorem.
Then $\overline{z}_{l}=\gamma \xi _{l}$ follows a Gaussian distribution with
0 mean and variance $\overline{\sigma }_{l}^{2}=\gamma ^{2}(\rho
\dsum\limits_{i\in \mathcal{I}_{l}^{c}}\psi _{l,i}^{2}+1)$.

Recall that $v_{l}\triangleq \mathbf{a}_{l}^{T}\otimes \mathbf{c}$. With the
above arrangement, the APP\ w.r.t. the $l$th ILC is now given by
\begin{equation}
p\left( v_{l}=\theta |\overline{y}_{l}\right) 
=%
\frac{1}{\eta }\sum_{\mathbf{c}:\mathbf{a}_{l}^{T}\otimes \mathbf{c=}\theta
}\exp \left( -\left\vert \overline{y}_{l}-\dsum\limits_{i\in \mathcal{I}_{l}}%
\sqrt{\rho }\psi _{l,i}c_{i}\right\vert ^{2}/2\overline{\sigma }%
_{l}^{2}\right) ,  \label{Eq_symbolwiseAPP_LinearFiltering}
\end{equation}%
where $\eta $ is the normalization factor. The APP $p\left( v_{l}=\theta |%
\overline{y}_{l}\right) $ is equal to the sum of the likelihood functions of
the candidates whose underlying ILC is equal to $\theta $.

\subsection{Low-complexity LCMA Soft Detection based on Gaussian
Approximation}

A direct execution of (\ref{Eq_symbolwiseAPP_LinearFiltering}) requires to
evaluate the Euclidean distances of $2^{m\omega \left( \mathbf{a}_{l}\right)
}$ candidates of $\mathbf{c}$. The order of complexity therein is thus $%
O(2^{m\omega \left( \mathbf{a}_{l}\right) })$. In this part, we develop
efficient soft ILC detection algorithms for the computation of (\ref%
{Eq_symbolwiseAPP_LinearFiltering}). The algorithms have per-user complexity
(much) less than $O(K)$.

In general, there is a many-to-one mapping between $\mathbf{a}_{l}^{T}%
\mathbf{c}$ and $\mathbf{a}_{l}^{T}\mathbf{\otimes c}$. Specifically, all
the events $\left\{ \mathbf{a}_{l}^{T}\mathbf{c=}\theta \pm \beta \cdot
2^{m}\right\} $ with various values $\overline{\theta }=\theta \pm \beta
\cdot 2^{m}$ have an identical $\mathbf{a}_{l}^{T}\mathbf{\otimes c=}\theta $
after the modulo-$2^{m}$ operation. As such, using the Total Probability
Rule, the APP is written as%
\begin{equation}
p\left( v_{l}=\theta |\overline{y}_{l}\right) =\frac{1}{\eta }\sum_{%
\overline{\theta }:\func{mod}\left( \overline{\theta },2^{m}\right) =\theta
\text{ }}p\left( \overline{y}_{l}|\mathbf{a}_{l}^{T}\mathbf{c}=\overline{%
\theta }\right) p\left( \overline{\theta }\right) .
\end{equation}

Here we derive the likelihood function $p\left( \overline{y}_{l}|\mathbf{a}%
_{l}^{T}\mathbf{c}=\overline{\theta }\right) $. Let $\Omega _{l}\left(
\overline{\theta }\right) =\left\{ \mathbf{c}:\mathbf{a}_{l}^{T}\mathbf{c}=%
\overline{\theta }\right\} $ collect the candidates $\mathbf{c}$ with $%
\mathbf{a}_{l}^{T}\mathbf{c}$ equal to $\overline{\theta }$. The conditional
mean for a given value of $\mathbf{a}_{l}^{T}\mathbf{c=}\overline{\theta }$
is
\begin{equation}
\mu _{l}\left( \overline{\theta }\right) 
=E_{\mathbf{c}}\left( \overline{y}%
_{l}|\mathbf{a}_{l}^{T}\mathbf{c}=\overline{\theta }\right) 
=%
\frac{1}{\left\vert \Omega _{l}\left( \overline{\theta }\right) \right\vert }%
\dsum\limits_{\mathbf{c\in }\Omega _{l}\left( \overline{\theta }\right)
}\dsum\limits_{i\in \mathcal{I}_{l}}\sqrt{\rho }\psi _{l,i}c_{i}.
\label{Eq_ConditionalMean}
\end{equation}%
The conditional variance is
\begin{eqnarray}
&\sigma _{l}^{2}\left( \overline{\theta }\right) = E_{\mathbf{c}}\left(
\left\vert \dsum\limits_{i\in \mathcal{I}_{l}}\sqrt{\rho }\psi _{l,i}c_{i}+%
\overline{z}_{l}-\mu _{l}\left( \overline{\theta }\right) \right\vert
^{2}\right) 
\notag \\
&= \frac{1}{\left\vert \Omega _{l}\left( \overline{\theta }\right)
\right\vert }\dsum\limits_{\mathbf{c\in }\Omega _{l}\left( \overline{\theta }%
\right) }\left( \dsum\limits_{i\in \mathcal{I}_{l}}\sqrt{\rho }\psi
_{l,i}c_{i}\right) ^{2}-\mu _{l}^{2}\left( \overline{\theta }\right) +\gamma
^{2}\widetilde{\sigma }_{l}^{2}.  \label{Eq_ConditionalVar}
\end{eqnarray}%
For a sufficiently large $K$, the proposed low-complexity algorithm
approximates $\overline{y}_{l}$ to have a conditional Gaussian distribution
for all values of $\overline{\theta }$. The APP is then calculated as
\begin{equation}
p\left( v_{l}=\theta |\overline{y}_{l}\right) =\frac{1}{\eta }\dsum\limits_{%
\overline{\theta }:\func{mod}\left( \overline{\theta },2^{m}\right) =\theta
}\exp \left( -\frac{\left( \overline{y}_{l}-\mu _{l}\left( \overline{\theta }%
\right) \right) ^{2}}{2\sigma _{l}^{2}\left( \overline{\theta }\right) }%
\right) p\left( \overline{\theta }\right) .  \label{Eq_APP}
\end{equation}%
This is referred to as \emph{Detection Method I} in this paper.

\subsubsection{Calculation of the Statistic Values in Detail}

Detection Method I requires a) the a priori probability $p\left( \overline{%
\theta }\right) $, b) the conditional mean $\mu _{l}\left( \overline{\theta }%
\right) $ and c) conditional variance $\sigma _{l}^{2}\left( \overline{%
\theta }\right) $ (\ref{Eq_APP}), to be detailed below. Since these
statistics are required to be calculated once per-block, the cost is minor
compared to that in (\ref{Eq_APP}) which are computed $n$ times per-block.
For notational simplify, the index $l$ is omitted in this part.

a) Let $n_{1}\left[ \overline{\theta }\right] =1$ for $\overline{\theta }%
=0,a_{1},\cdots ,(2^{m}-1)a_{1}$ if $a_{1}>0$, and $\overline{\theta }%
=(2^{m}-1)a_{1},\cdots ,0$ if $a_{1}<0$. Let $n_{1}\left[ \overline{\theta }%
\right] =0$ for the rest values of $\overline{\theta }$. Then $p\left(
\overline{\theta }\right) $ can be obtained by sequentially implementing%
\begin{equation}
n_{k}\left[ \overline{\theta }\right] =\sum_{\tau =0,\cdots ,2^{m}-1}n_{k-1}%
\left[ \overline{\theta }-a_{i}\tau \right]  \label{Eq_APR}
\end{equation}%
until layer $K^{\prime }=\omega \left( \mathbf{a}\right) $ is reached. This
requires no more than $\sum_{k=1}^{K^{\prime }}\left( \omega _{H}\left( \left[ a_{1},\cdots ,a_{k}%
\right] \right) (2^{m}-1)+1\right) \left( 2^{m}-1\right) \approx
\sum_{k=1}^{K^{\prime }}\omega _{H}\left( \left[ a_{1},\cdots ,a_{k}\right]
\right) (2^{m}-1)^{2}$
additions in total and does not involve multiplication.

b) The conditional means can be obtained by sequentially implementing
\begin{equation}
\widetilde{\mu }_{k}\left[ \overline{\theta }\right] =\sum_{\tau =0,\cdots
,2^{m}-1}\widetilde{\mu }_{k-1}\left[ \overline{\theta }-a_{i}\tau \right]
+\tau \sqrt{\rho }\psi _{k}.
\end{equation}%
When reaching layer $K^{\prime }=\omega \left( \mathbf{a}\right) $ , the
conditional mean is computed by $\mu \left( \overline{\theta }\right) =%
\widetilde{\mu }_{K^{\prime }}\left[ \overline{\theta }\right] /n_{K^{\prime
}}\left[ \overline{\theta }\right] $. 

c) The term $\dsum\limits_{\mathbf{c\in }\Omega \left( \overline{\theta }%
\right) }\left( \dsum\limits_{i\in \mathcal{I}}\sqrt{\rho }\psi
_{i}c_{i}\right) ^{2}$ is calculated by sequentially implementing $\vartheta _{k}\left[ \overline{\theta }\right] =$
\begin{align}
\sum_{\tau =0,\cdots
,2^{m}-1}\left( \vartheta _{k-1}\left[ \overline{\theta }-a_{k}\tau \right]
+2\tau \sqrt{\rho }\psi _{i}u_{k-1}\left[ \overline{\theta }-a_{k}\tau %
\right] +\left( \tau \sqrt{\rho }\psi _{i}\right) ^{2}\right) .  \notag
\end{align}
When reaching layer $K^{\prime }$, the conditional variance is obtained as
\begin{equation}
\sigma ^{2}\left( \overline{\theta }\right) =s_{K^{\prime }}\left[ \overline{%
\theta }\right] /n_{K^{\prime }}\left[ \overline{\theta }\right] -\mu
^{2}\left( \overline{\theta }\right) +\gamma ^{2}\overline{\sigma }^{2}.
\end{equation}

\subsubsection{Complexity Analysis}

Here, it can be easily shown that the integer-valued $\overline{\theta }$
are within the range of
\begin{equation}
\overline{\theta }\in \{\dsum\limits_{i:a_{l,i}<0}a_{l,i}\left(
2^{m}-1\right) ,\cdots ,\dsum\limits_{i:a_{l,i}>0}a_{l,i}\left(
2^{m}-1\right) \}.
\end{equation}%
Define $\omega _{H}\left( \mathbf{a}_{l}\right) \triangleq
\dsum\limits_{i\in \mathcal{I}_{l}}\left\vert a_{l,i}\right\vert ,$ referred
to as the \textquotedblleft \emph{weight\textquotedblright } of $\mathbf{a}%
_{l}$. Then the cardinality of the set for $\overline{\theta }$ is precisely
$\omega _{H}\left( \mathbf{a}_{l}\right) (2^{m}-1)+1$. In other words, there
are $\omega _{H}\left( \mathbf{a}_{l}\right) (2^{m}-1)+1$ Euclidean
distances needs to be calculated in (\ref{Eq_APP}). This is far less than $%
2^{m\omega \left( \mathbf{a}_{l}\right) }$ required in direct execution of (%
\ref{Eq_symbolwiseAPP_LinearFiltering}).

For all $K$ ILCs, the total number of Euclidean distance calculations is%
\begin{equation}
(2^{m}-1)\dsum\limits_{l=1}^{K}\omega _{H}\left( \mathbf{a}_{l}\right)
+K
=2^{m}K\cdot E_{\mathbf{a}}\left( \omega _{H}\left( \mathbf{a}\right) \right)
\end{equation}%
where $E_{\mathbf{a}}\left( \omega _{H}\left( \mathbf{a}\right) \right) $ is
the average weight of coefficient vectors. The average per-user complexity
has order $O\left( 2^{m}E_{\mathbf{a}}\left( \omega _{H}\left( \mathbf{a}%
\right) \right) \right) $. This is $E\left( \omega _{H}\left( \mathbf{a}%
\right) \right) $ times of the complexity of single-user detection. As we
will see later in Section VI, $E_{\mathbf{a}}\left( \omega _{H}\left(
\mathbf{a}\right) \right) $ is just a fraction of $K$ in general for $K$
being large.

\subsection{Decoding}

The soft ILC detection outcome $p\left( v_{l}\left[ t\right] |\mathbf{y}%
\left[ t\right] \right) ,t=1,\cdots ,n$ is forwarded to a ring code decoder,
which carried out $2^{m}$-ary decoding algorithm to obtain its soft outputs
on the message level ILC $u_{l}\left[ t\right] ,t=1,\cdots ,k$. For a $2^{m}$%
-ary LDPC ring code \cite{YuTcom22}, for example, $2^{m}$-ary belief
propagation (BP) algorithm is employed. We again note that the soft
detection and decoding for the $K$ ILC streams are executed in parallel.
Upon all $K$ message-level ILCs are calculated, the messages of the $K$
users are recovered by (\ref{Eq_RecoverMessages}).

\subsection{Example with Integer-forcing}

Our developed algorithm applies to any filter matrix $\mathbf{W=}\left[
\mathbf{w}_{1},...,\mathbf{w}_{K}\right] ^{T}$. If exact IF (EIF) is adopted
in a $K\leq N_{S}$ system, the filter matrix is given by \cite{ZhanIT14} $%
\mathbf{W}_{IF}\mathbf{=A}\left( \mathbf{S}^{T}\mathbf{\mathbf{S}}\right)
^{-1}\mathbf{S}^{T}.$ The signal is then given as
\begin{equation}
\overline{y}_{l}=\dsum\limits_{i\in \mathcal{I}_{l}}\sqrt{\rho }\psi
_{l,i}c_{i}+\overline{z}_{l}=\dsum\limits_{i\in \mathcal{I}_{l}}\sqrt{\rho }%
a_{l,i}c_{i}+\overline{z}_{l}.
\end{equation}%
The last equality follows from the fact that $\left( \mathbf{S}^{T}\mathbf{%
\mathbf{S}}\right) ^{-1}\mathbf{S}^{T}\mathbf{S=I.}$ In this case, the
effective gains are exactly identical to the coefficient vectors. Thus, $\mu
_{l}\left( \overline{\theta }\right) =\overline{\theta }$ and $\sigma
_{l}^{2}\left( \overline{\theta }\right) =\gamma ^{2}\widetilde{\sigma }%
_{l}^{2}$, which are utilized in (\ref{Eq_APP}) to calculate the APP of ILC.
The a priori probabilities are required to be calculated as in (\ref{Eq_APR}%
) .

The EIF presented above does not support an MA setup of $K>N_{S}$, and
suffers from performance loss, particularly at low SNR. Regularized IF (RIF)
can address such issues, whose filter matrix is \cite{ZhanIT14}
\begin{equation}
\mathbf{W}_{RIF}\mathbf{=AS}^{T}\left( \rho \mathbf{S\mathbf{S}}^{T}\mathbf{+%
}I_{N}\right) ^{-1}.  \label{Eq_RIF}
\end{equation}%
With $\mathbf{W}_{RIF}$, the signal can be written as%
\begin{equation}
\overline{y}_{l}=\dsum\limits_{i\in \mathcal{I}_{l}}\sqrt{\rho }\psi
_{l,i}c_{i}+\overline{z}_{l}=\dsum\limits_{i\in \mathcal{I}_{l}}\sqrt{\rho }%
a_{l,i}c_{i}+e_{l}.  \label{Eq_SignalafterRIF}
\end{equation}%
The estimation error term is
\begin{equation}
e_{l}=\dsum\limits_{i\in \mathcal{I}_{l}}\sqrt{\rho }\left( \psi
_{l,i}-a_{l,i}\right) c_{i}+\overline{z}_{l}\mathbf{.}  \label{Eq_e}
\end{equation}%
Unlike the EIF, here the error term $e_{l}$ in RIF is correlated with the
useful signal part $\dsum\limits_{i\in \mathcal{I}_{l}}\sqrt{\rho }%
a_{l,i}c_{i}$. This leads to $\mu _{l}\left( \overline{\theta }\right) \neq
\overline{\theta }$ and $\sigma _{l}^{2}\left( \overline{\theta }\right)
\neq \gamma ^{2}\widetilde{\sigma }_{l}^{2}$, which must be calculated as in
(\ref{Eq_ConditionalMean}) and (\ref{Eq_ConditionalVar}), respectively.

For a sufficiently large $K$, the number of terms that adds up in (\ref{Eq_e}%
) is sufficiently large to apply Central Limit Theorem for $e_{l}$. Hence,
one may approximate $e_{l}$ as a Gaussian random variable with variance $%
E\left( e_{l}^{2}\right) $. It can be\ easily shown that the MSE of $e_{l}$
has a closed-form representation
\begin{equation}
E\left( e_{l}^{2}\right) =\gamma ^{2}\mathbf{a}_{l}^{T}\left( \rho \mathbf{S}%
^{T}\mathbf{S+I}\right) ^{-1}\mathbf{a}_{l}^{T}.  \label{Eq_MSERIF}
\end{equation}%
Further, by disregarding the bias in the estimation error term, the mean of $%
e_{l}$ is approximated as zero. As such, the calculation of the APP in (\ref%
{Eq_APP}) is further simplified into
\begin{equation}
p\left( v_{l}=\theta |\overline{y}_{l}\right) \approx \frac{1}{\eta }%
\dsum\limits_{\overline{\theta }:\mathbf{a}_{l}^{T}\otimes \mathbf{c}=\theta
}\exp \left( -\frac{\left( \overline{y}_{l}-\overline{\theta }\right) ^{2}}{%
2E\left( e_{l}^{2}\right) }\right) p\left( \overline{\theta }\right) .
\label{Eq_APPRIF}
\end{equation}%
This is referred to as \emph{Detection Method II}, which is inferior to
Detection Method I due to the approximation in the conditional mean and
variance. Note that the a priori probabilities are required to be calculated
as in (\ref{Eq_APR}), while the calculations of conditional means and
variances are avoided.

For a relatively small number of users $K$, Detection Method I is suggested
while the loss of Method II may be considerable. For a large value of $K$,
either Method I or Method II could be used. The receiver processing with
linear LCMA detection method II is summarized in Algorithm 1.

\begin{algorithm}
\caption{Summary of LCMA Receiver's Processing (Detection Method II.) }
Step 1) Calculate the filtering matrix $\mathbf{W}_{RIF}$ according
to (\ref{Eq_RIF}). Perform the filtering process (\ref{Eq_SignalafterRIF}).

Step 2) Calculate the MSE $E\left( e_{l}^{2}\right) $ according to (\ref%
{Eq_MSERIF}) for $l=1,\cdots,K$. Calculate the a priori probabilities as in (\ref{Eq_APR}).

Step 3) Perform (\ref{Eq_APPRIF}) in parallel to calculate the APPs $p\left(
v_{l}=\theta |\overline{y}_{l}\right) $ for the $K$ streams of
codeword-level ILCs. Forward the $K$ streams of APPs to the $K$ ring-code
decoders.

Step 4) Perform ring-code decoding for the $K$ streams parallely, which
yields the decisions on $\mathbf{u}_{1},..,\mathbf{u}%
_{K}$.

Step 5) Recover $K$ users' messages by implementing $\mathbf{B=}\widetilde{%
\mathbf{A}}^{-1}\mathbf{\otimes U}$.
\end{algorithm}
 {\subsection{Treatment for MU-MIMO}}
 {The LCMA technique and algorithms developed above directly apply to MU-MIMO,
by simply replacing the spreading matrix $\mathbf{S}$ by the MIMO channel
coefficient matrix $\mathbf{H}$ in executing the LCMA receiver processing. A
slight difference is that one may control/optimize the spreading matrix $%
\mathbf{S}$ for the G-MAC, while the MIMO channel coefficient matrix $%
\mathbf{H}$ depends wholly on the fading channel realization which cannot be
controlled. In addition, the conversion from a $K$-by-$N$ complex-valued model to a $2K$-by-$2N$ real-value model in (3) enables us to directly utilize the (real-value based) LCMA algorithms developed above.}
\section{On the Optimized Design of LCMA}

\subsection{Design of ILC Coefficient Matrix $\mathbf{A}$}

In this section, we focus on LCMA with linear detection and provide a
efficient yet powerful pragmatic solution to\footnote{%
The optimized $\mathbf{A}$ here is not optimal for the non-linear soft
detectors \cite{YangTaoCL23}, but also works reasonably well therein.} $\mathbf{%
A}$, for a given spreading matrix $\mathbf{S}$. Following the convention in
studying uplink MA, we consider that all users have symmetric rate. Our
development can be extended to non-symmetric rates. For a given spreading
matrix $\mathbf{S}$, let the minimum mean square error (MMSE) matrix be
denoted by $\mathbf{\Psi =}\left( \rho \mathbf{\mathbf{S}}^{T}\mathbf{S+I}%
_{K}\right) ^{-1}$. Its eigen-decomposition is%
\begin{equation}
\mathbf{\Psi =VDV}^{T}.
\end{equation}%
The rate for reliable communication of LCMA is characterized in the
following theorem.

\begin{theo}
A symmetric rate $R_{0}$ is achievable if there exists $K$ integer vectors $%
\mathbf{a}_{1},\cdots ,\mathbf{a}_{K}$, that are linearly independent in $%
\mathbb{Z}_{2^{m}}$, such that
\begin{equation}
\mathbf{D}^{\frac{1}{2}}\mathbf{V}^{T}\mathbf{a}_{l}<\sqrt{\frac{1}{%
2^{2R_{0}}}},\forall l=1,\cdots ,K.  \label{Eq_theorem1}
\end{equation}
\end{theo}

\begin{proof}[Proof of Theorem 1]
: It can be shown that the MSE in the linear estimator of $\mathbf{a}_{l}^{T}%
\mathbf{x}\left[ t\right] $ is $\min_{\mathbf{w}_{l}}E\left( \left\vert \mathbf{w}_{l}^{T}\mathbf{y}\left[ t%
\right] \mathbf{-a}_{l}^{T}\mathbf{x}\left[ t\right] \right\vert ^{2}\right)$
\begin{equation}
=\mathbf{a}_{l}^{T}\left( \rho \mathbf{\mathbf{S}}^{T}\mathbf{S+I}%
_{K}\right) ^{-1}\mathbf{a}_{l}\mathbf{=a}_{l}^{T}\mathbf{VDV}^{T}\mathbf{a}%
_{l}.
\end{equation}%
As $n$ tends to infinity, the effective noise sphere is given by this MSE
value for computing the $l$th ILC. There exist a nested lattice-code with
simultaneous \textquotedblleft Roger-goodness\textquotedblright\ and
\textquotedblleft Poltyrev-goodness\textquotedblright , such that the rate
\begin{equation}
R_{l}^{comp}=\frac{1}{2}\log _{2}^{+}\left( \frac{1}{\mathbf{a}_{l}^{T}%
\mathbf{VDV}_{l}^{T}\mathbf{a}}\right) .  \label{Eq_RateComp}
\end{equation}%
w.r.t. the $l$th ILC is achievable \cite{NazerIT11,ZhanIT14}. The overall
achievable symmetric rate is given by $R_{0}\leq R_{sym}$ with
\begin{equation*}
R_{sym}=\min_{l=1,\cdots ,K}R_{l}^{comp}=\min_{l=1,\cdots ,K}\frac{%
1}{2}\log _{2}^{+}\left( \frac{1}{\mathbf{a}_{l}^{T}\mathbf{VDV}_{l}^{T}%
\mathbf{a}}\right) .
\end{equation*}%
Then, all $K$ ILCs can be reliably computed if
\begin{equation}
\min_{l=1,\cdots ,K}\frac{1}{2}\log _{2}^{+}\left( \frac{1}{\mathbf{a}%
_{l}^{T}\mathbf{VDV}_{l}^{T}\mathbf{a}}\right) >R_{0},  \label{Eq_RateTarget}
\end{equation}%
then all users' messages can be recovered. As (\ref{Eq_RateTarget}) is
equivalent to (\ref{Eq_theorem1}), the proof is completed.
\end{proof}

\begin{remark}
The above characterization is based on the existence of nested lattices that
are simultaneously good for shaping and channel-coding. For lattice-based
processing with the practical ring-coded PAM, the achievable mutual
information that takes into account the $2^{m}$-PAM should be used for
characterization. However, for the case with a large number of UEs and BS
antennas, it is well-known that calculating the exact mutual information for
$2^{m}$-PAM requires a multi-dimension integration. This makes finding the
optimized $\mathbf{A}$ intangible. To obtain a viable and pragmatic solution
to $\mathbf{A}$, in this paper we resort to the succinct rate expression (%
\ref{Eq_RateComp}) in our treatment. Optimization over such characterization
yields a competitive solution for our developed LCMA with ring-coded PAM, as
we will see in the next section.
\end{remark}

Given Theorem 1, the problem is now to find $K$ linearly independent lattice
points, formed by the basis $\mathbf{D}^{\frac{1}{2}}\mathbf{V}^{T}$, within
the boundary of radius $\sqrt{\frac{1}{2^{2R_{0}}}}$. This is referred to as
a \emph{bounded independent vectors problem }(BIVP). Solving the BIVP is
easier than solving the shortest independent vector problem (SIVP), as one
only need $K$ independent points within the radius $\sqrt{\frac{1}{2^{2R_{0}}%
}}$ rather than the $K$ shortest ones. The linear independence of $\mathbf{a}%
_{1},\cdots ,\mathbf{a}_{K}$ is w.r.t. $\mathbb{Z}_{2^{m}}$, which
guarantees that $\widetilde{\mathbf{A}}=\func{mod}\left( \mathbf{A}%
,2^{m}\right) $ has a unique inverse in $\mathbb{Z}_{2^{m}}$. For a
relatively large $K$ and $m$, the linear independence w.r.t. $\mathbb{Z}%
_{2^{m}}$ is equivalent to that in $\mathbb{Z}$ in probability.

The most well-known approach, that may be borrowed to address the SIVP
problem, would be the Lenstra--Lenstra--Lov\'{a}sz (LLL) algorithm \cite%
{LenstraLLL}. The basic idea of the LLL algorithm is that, for a given set
of lattice basis vectors, one carries out integer linear combinations of
them to obtain new shorter lattice vectors. This step exploits the mechanism
of Gram-Schmidt orthogonalization, with the weights being integers. The
resultant vectors are then re-arranged (exchanged) in an increasing order of
length, and the next round of integer linear combinations of the vectors are
performed. In particular, the size-reduce condition and the Lov\'{a}sz
condition are employed to guide the integer linear combinations and the
exchanges of the lattice vectors. It is shown that LLL algorithm has a
complexity order that is polynomial in $K$. However, for a large $K$, the
performance loss of LLL becomes dramatic. The literature of lattice
reduction is rich in computer science area. Other well-known approaches for
the SVP and SIVP include LLL-boost\cite{LyuTSP17}, HKZ \cite{SakzadTWC13},
BKZ and BKZ 2.0 and etc.

In fact, we are not indeed solving the SIVP, but just the BIVP. Owing to
this, we now propose a \emph{rank-constrained sphere decoding} (RC-SD)
algorithm which solves the BIVP in (\ref{Eq_theorem1}). The goal is to find $%
K$ coefficient vectors $\mathbf{a}_{1},\cdots ,\mathbf{a}_{K}$ that are 1)
within the boundary of radius $\sqrt{\frac{1}{2^{2R_{0}}}}$ and 2) has
full-rank $K$. In RC-SD, we start with an initial radius that is smaller
than $\sqrt{\frac{1}{2^{2R_{0}}}}$ and apply a sphere decoding tree search.
If the rank for the candidates within the boundary is smaller than $K$, the
radius is added by a certain small step and continue the tree search, until
the rank reaches $K$. Then, we pick those $K$ linearly independent vectors
with smallest norms. The pseudo code for this approach is given in Algorithm
2.

\begin{algorithm}[h]
\caption{Rank-constrained Sphere Decoding for identifying $\mathbf{A}$}
Step 1) Set the radius to $r=\sqrt{\frac{1}{2^{2R_{0}}}}-\varepsilon $. The
initial value can be set empirically.

Step 2) Implement the Cholesky factorization on $\mathbf{VDV}_{l}^{T}$, and
apply a tree-search method which finds all candidates $\mathbf{a}$ that are
within distance $r$ to the origin.

Step 3) If the rank of the candidates in $\mathbb{Z}_{2^{m}}$ is less than $K
$. Increase the radius $r$ by a certain (small) step and Go to Step 1).
Otherwise, continue to Step 4).

Step 4)  Arrange the candidates with a ascending order according to the
lengths. Start with the first candidate, use a greedy method to find $K
$ coefficient vectors, until the rank in reaches $K$.

\end{algorithm}


\begin{figure}[h]
\centering
\includegraphics[width=0.4\textwidth]{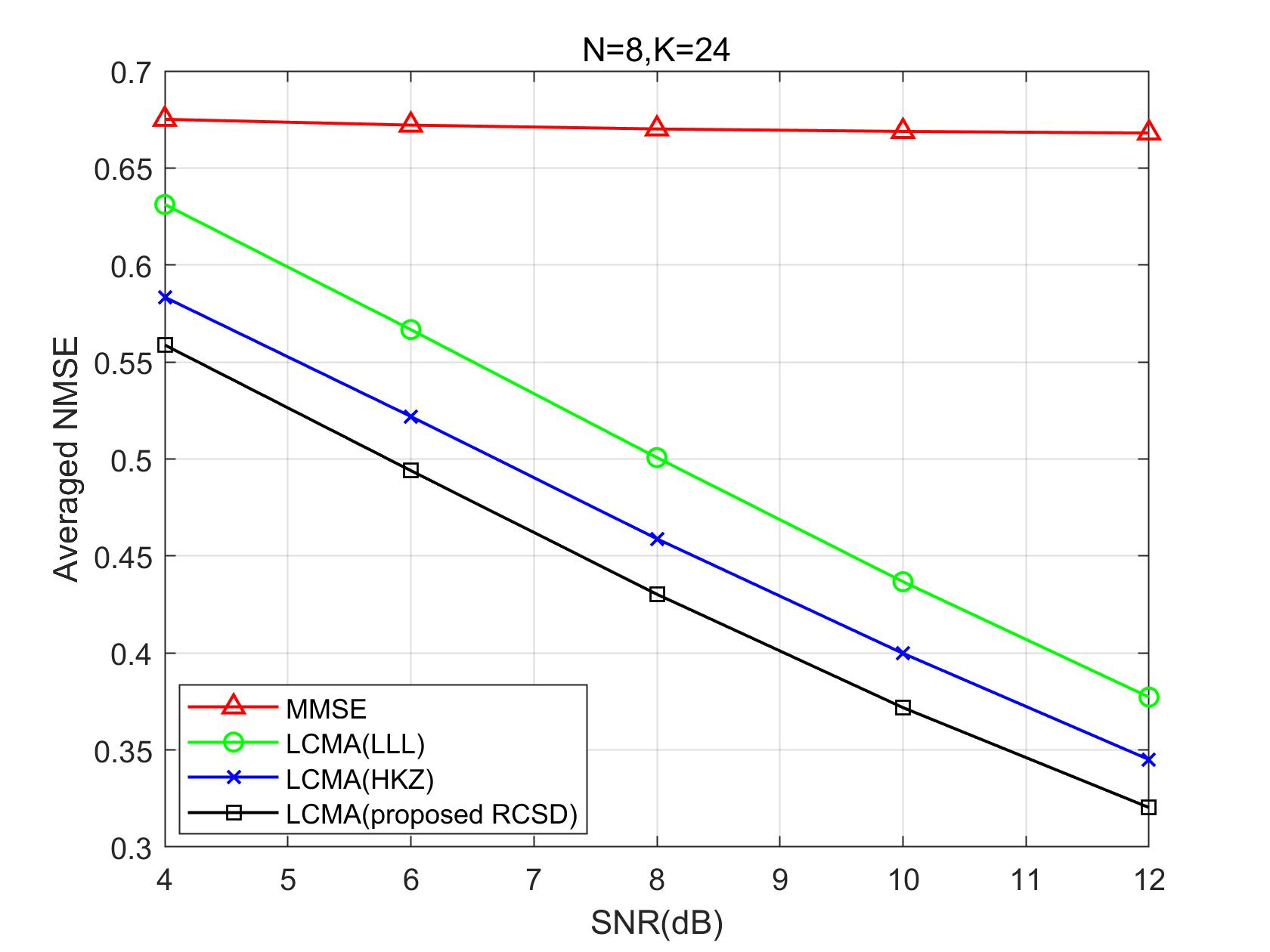}
\caption{Averages NMSE with the proposed RS-SD, $N=8,K=24$. The channel
coefficients follows Rayleigh distribution.}
\label{NMSE_RCSDvsLLL}
\end{figure}

Fig. \ref{NMSE_RCSDvsLLL} shows the averaged normalized MSE (NMSE) w.r.t.
the ILCs where $N=8,K=24$, i.e, the system load of LCMA is 300\%. Here $%
\mathbf{\mathbf{S}}$ is randomly given. It is well-known that NMSE
characterizes the quality of the soft detection output. The smaller the
NMSE, the greater the mutual information or supported code rate, following
the notion of sphere-packing \cite{ErezTIT04}. The conventional
non-lattice-coded based scheme with MMSE detection fails to support this
system load, as the MSE barely drops as SNR increases. In contrast, LCMA
with any of the three methods for obtaining coefficient matrix $%
\mathbf{A}$ can support a system load of 300\%. In particular, our proposed
RC-SD method considerably outperforms existing LLL and HKZ lattice reduction
methods \cite{SakzadTWC13}. We note that the efficiency of RC-SD algorithm
may be further improved by jointly considering the full-rank condition in
reducing the dimension of the search space, which will be investigated in
future work.

Notably, LCMA is different from lattice-reduction based MIMO detection \cite%
{SakzadTWC13}. LCMA utilizes an $n$-dimension ring-coded PAM as the
underlying coding-modulation, where the lattice is characterized by the
generator matrix $\mathbf{G}$. The optimization of LCMA is based on the
lattice obtained from the MMSE matrix. Lattice-reduction based MIMO
detection is dealing with the lattice generated by the channel matrix $%
\mathbf{H}$. Our work differs from the lattice partition multiple access
scheme \cite{FangGlobecom16} in the following. First, \cite{FangGlobecom16}
considered a closed-loop downlink scenario and utilized superposition of the
multiple users' signals via lattice partitioning. Only the theoretical rate
was studied while no practical lattice codes were studied therein. In
contrast, our work considers an open-loop uplink system, and is based on the
property of the underlying lattice code (in the form of a practical $2$-ary
ring coded PAM). The dramatic improvement in terms of system load and BLER
is presented. Second, the notions of structured binning and the computation
of integer linear combinations, exploited in our work, are not relevant to
the work in \cite{FangGlobecom16}. Furthermore, \cite{FangGlobecom16} was
based on the successive cancellation procedures at the UE nodes, while in
our work parallel processing procedures are utilized at the BSs. New
algorithms for identifying the optimized $\mathbf{A}$ matrix and the
spreading matrix are developed in our work, which are not studied in \cite%
{FangGlobecom16}.

\subsection{Design of LCMA Signature Sequences $\mathbf{S}$}
So far we have presented the algorithms of the LCMA receiver: the soft ILC
detection and the identification of $\mathbf{A}$. We are now in the position
to study the spreading signature sequences $\mathbf{S}=[\mathbf{s}%
_{1},\cdots ,\mathbf{s}_{K}]$ for LCMA. Obviously, the
performance of LCMA is a function of both $\mathbf{S}$ and $\mathbf{A}$, and
the joint optimized design problem is formulated as
\begin{eqnarray}
&&\arg \max_{\mathbf{S,A}}\underset{l}{\text{min}}\frac{1}{2}\log
_{2}^{+}\left( \frac{1}{\mathbf{a}_{l}^{T}\left( \rho \mathbf{\mathbf{S}}^{T}%
\mathbf{S+I}_{K}\right) ^{-1}\mathbf{a}_{l}}\right)
\label{Eq_JointOptFormulation} \\
&&\text{s.t. \ Tr}\left\{ \mathbf{S}^{T}\mathbf{S}\right\} \leq K\text{, }%
\mathbf{A}\in\mathbb{Z}^{K\times K}\text{ has full rank }K,  \notag
\end{eqnarray}%
where a total power constraint is considered. The formulation w.r.t.
individual power constraint just needs to modified the power constraint into
the form $\left\Vert \mathbf{s}_{i}\right\Vert ^{2}\leq 1,\forall i=1,\cdots
,K$.

In this paper, we suggest a
pragmatic method that decouples the optimization of $\mathbf{S}$ and $%
\mathbf{A}$ as follows. First, we solve the optimization of $\mathbf{S}$ for
a given $\mathbf{A}$. A local optima to the problem in (\ref%
{Eq_JointOptFormulation}) can be approximately found using a
target-switching steepest descent (TS-SD) algorithm. The main idea of the TS-SD method is that the steepest descent based optimization process
always targets on the ILC stream with the lowest compute rate, i.e., \begin{equation}l_{min}=\arg \min_l{R}%
_{l}^{Comp}\label{Eq_Lmin}\end{equation} where ${R}
_{l}^{Comp}$ is given in (\ref{Eq_RateComp}). For
the $l_{min}$-th ILC, we establish a Lagrangian function as 
\begin{equation}
\varphi (\mathbf{S},u)=\mathbf{a}_{l_{min}}^{T}{(\rho \mathbf{S}^{T}%
\mathbf{S}+\mathbf{I}_{K})}^{-1}\mathbf{a}_{l_{min}}+u(\text{tr}(\mathbf{%
S}^{T}\mathbf{S})-K).
\end{equation}

The steepest descent direction is obtained by taking the partial derivatives
of $\varphi (\mathbf{S},u)$ w.r.t. $\mathbf{S}$ and $u$, given by 
{\small
\begin{align}
& \frac{\partial \varphi (\mathbf{S},u)}{\partial \mathbf{S}}=2{u}\mathbf{S}%
-2\rho \mathbf{S}{(\rho \mathbf{S}^{T}\mathbf{S}+\mathbf{I}_{K})}^{-1}%
\mathbf{a}_{l_{min}}\mathbf{a}_{l_{min}}^{T}{(\rho \mathbf{S}^{T}\mathbf{S}+\mathbf{I}%
_{K})}^{-1}, \\
& \frac{\partial \varphi (\mathbf{S},u)}{\partial u}=\text{tr}(\mathbf{S}^{T}%
\mathbf{S})-K.
\end{align}%
}
Then, $\mathbf{S}$ is updated by 
\begin{equation}
\mathbf{S}=\mathbf{S}-{\alpha }(2{u}\mathbf{S}-2\rho \mathbf{S}{(\rho {%
\mathbf{S}}^{T}\mathbf{S}+\mathbf{I}_{K})}^{-1}\mathbf{a}_{l_{min}}\mathbf{a}%
_{l_{min}}^{T}{(\rho {\mathbf{S}}^{T}\mathbf{S}+\mathbf{I}_{K})}^{-1})
\end{equation}%
where $\alpha $ is the update step (or learning rate). We
consider that $\alpha $ is a sampling value of a hyperbolic tangent
function. Note that the initial value of ${u}$ should be made larger than $%
\alpha $ for the stability purpose of the iterative process. Also, ${u}$ is
updated by 
\begin{equation}
{u}={u}+{\alpha }(\text{tr}({\mathbf{S}}^{T}\mathbf{S})-K).
\end{equation}%
The update continues in an iterative manner as in standard SD.
%
%
%
pseudo code for the SD step is presented in Algorithm 3.

Next, for the given $\mathbf{S}$, solve the optimization of $\mathbf{A}$ as
in (\ref{Eq_theorem1}) with the RC-SD algorithm. Then, carry out iterations
between the above two steps until convergence is achieved.

As the optimization of spreading sequences $\mathbf{S}$ is off-line, one
may assign various initial values of $\mathbf{S}$ (or $\mathbf{A}$) and
select the one with the best cost function to approximate the global optima.

%
%

\begin{algorithm}[h]
	\begin{algorithmic}
		
		\renewcommand{\algorithmicrequire}{\textbf{Input:}}
		\renewcommand{\algorithmicensure}{\textbf{Output:}}
		
		\REQUIRE  $N_S,K,\rho,\alpha,u^{(0)}$
		\ENSURE $\mathbf{S}_{opt}$
		
		\STATE Initialization: $k=0$, $\mathbf{S}^{(0)}= $randn($N,K$),
		$\mathbf{S}^{(1)}=\frac{\mathbf{S}}{\sqrt{\text{tr}(\mathbf{S}^{T}\mathbf{S})/K}}$
				
		\WHILE{$\mathbf{S}^{(k+1)}\neq\mathbf{S}^{(k)}$}
		
		\STATE $k\leftarrow k+1$
		
		\STATE $l_{min}\leftarrow\min(R_1^{Comp},\cdots,R_K^{Comp}$)
		
		\STATE $\mathbf{S}^{(k+1)}\leftarrow\mathbf{S}^{(k)}-{\alpha} (2{u}^{(k)}\mathbf{S}^{(k)}-2\rho\mathbf{S}^{(k)}$\\
		${(\rho{\mathbf{S}^{(k)}}^{T}\mathbf{S}^{(k)}+\mathbf{I}_K)}^{-1}\mathbf{a}_{l_{min}}\mathbf{a}_{l_{min}}^{T}{(\rho{\mathbf{S}^{(k)}}^{T}\mathbf{S}^{(k)}+\mathbf{I}_K)}^{-1}) \notag$
		
		\STATE ${u}^{(k+1)}\leftarrow{u}^{(k)}+{\alpha}(\text{tr}({\mathbf{S}^{(k+1)}}^{T}\mathbf{S}^{(k+1)})-K)$
		
		\ENDWHILE
		
		\STATE $\mathbf{S}_{opt}=\mathbf{S}^{(k)}$
		
	\end{algorithmic}
	
	\caption{Target-switiching Steepest Descent}
	
	\label{Alg_TSSD}
\end{algorithm}

\begin{figure}[h]
\centering
\includegraphics[width=0.4\textwidth]{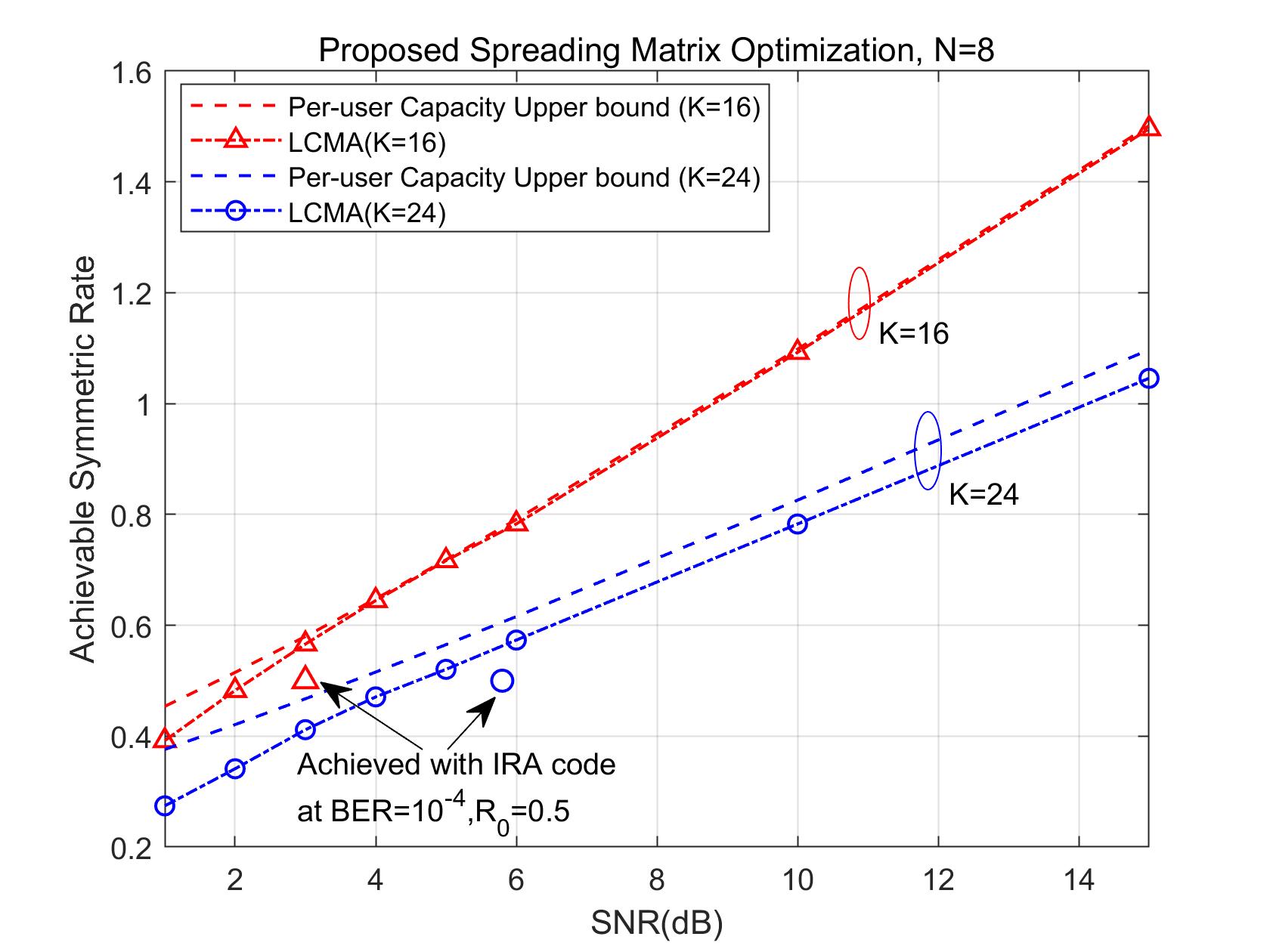}
\caption{Achievable symmetric rate per-user of LCMA with our designed
spreading matrix $\mathbf{S}$ in Gaussian MA channel, where $N_S=8,K=$16 and
24. Total power constraint is utilized here. Note that the horizontal axis
denotes the per-user SNR, while the vertical axis denotes the per-user rate.
The sum-rate is equal to $K$ times the per-user rate. }
\label{Fig_AchievableSymmetricRate}
\end{figure}

Fig. \ref{Fig_AchievableSymmetricRate} presents the achievable
rate of LCMA in an Gaussian MA channel
using our developed TS-SD method. Here we present with $N_S=8$, $K=16,24$. For SNR greater than 4 dB,
the difference between the achievable symmetric rate of LCMA and that of the
upper bound (UB) of the MA channel capacity is almost unnoticeable for $K$%
=16, and is quite small (about 0.05 bit) for $K$=24. At low SNRs, the gap
becomes greater. This is due to the well-known inherent loss of the
lattice-code based scheme that achieves $\frac{1}{2}\log ^{+}\left(
\varkappa +SNR\right) $, with $\varkappa <1$ in general \cite%
{NamIT10,NazerIT11}. We also include the
performance with a long IRA code. For $K=16$, at BER of $10^{-4}$, the
performance is about 0.71 dB and 1.04 dB away from the rate limit of LCMA
and MA capacity upper bound, respectively.



\section{Numerical Results}


\subsection{LCMA for Gaussian MA Channel}

Fig. \ref{Fig_AchievableSymmetricRate} also includes the performance of LCMA
for Gaussian MA channel with a capacity approaching irregular repeat
accumulate (IRA) code. The rate 1/2 binary irregular repeat accumulate (IRA)
code of length $k=65536$ is from that in \cite{ten2003design} optimized for
single-user point-to-point AWGN channel. The utilization of a long IRA
code is for the sake of comparison to the capacity limit. We observe
that for $K=16$, at BER of $10^{-4}$, the performance is about 0.71 dB and
1.04 dB away from the rate limit of LCMA and MA capacity UB, respectively.
For $K=24$, the performance is about 1.09 dB and 1.79 dB away from the rate
and capacity limits. The performance behavior of LCMA with a practical IRA
code is in line with the theoretical achievable rate result shown in Fig. %
\ref{Fig_AchievableSymmetricRate}.

\begin{figure}[h]
\centering%
\includegraphics[width=0.4\textwidth]{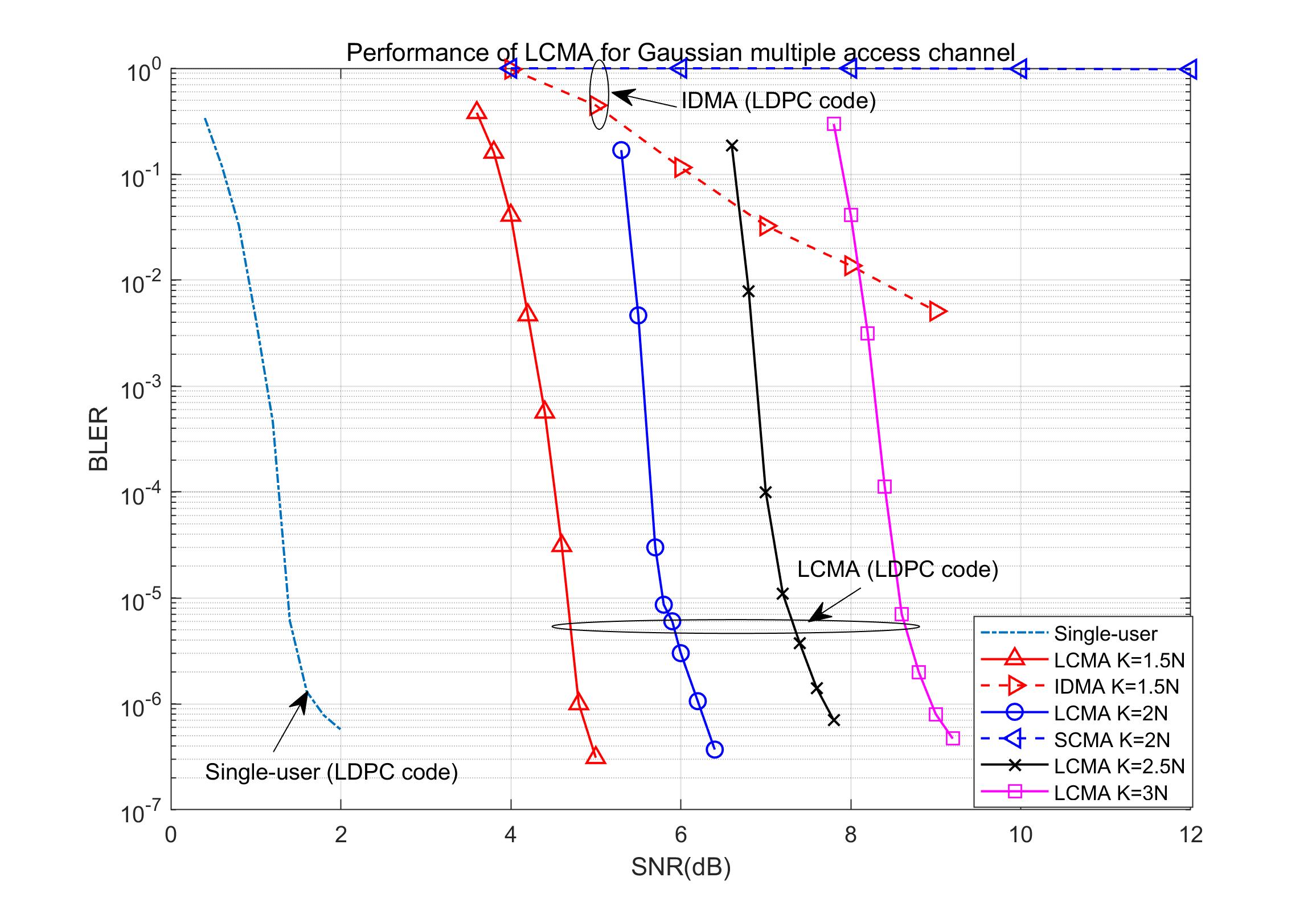}
\caption{BER of LCMA in Gaussian MA channel with $N_{s}=8$. BPSK and a rate
1/2 LDPC code of length $k=3840$ are utilized. The individual power
constraint is utilized in designing the spreading matrix $\mathbf{S}$.}
\label{Fig_FER_LCMA_LDPC_AWGNMA}
\end{figure}

\begin{table*}[h]
\caption{Example of spreading sequence of LCMA, $K$=12, $N_{S}$=8.}
\label{Table_SpreadingSequenceK8}\centering{\tiny
\begin{tabular}{|l|l|l|l|l|l|l|l|l|l|l|l|}
\hline
-0.4150 & 0.4828 & -0.0700 & -0.4150 & 0.4828 & -0.0700 & -0.4150 & 0.4828 &
-0.0700 & -0.4150 & 0.4828 & -0.0700 \\ \hline
-0.4150 & 0.4828 & -0.0700 & 0.4150 & -0.4828 & 0.0700 & -0.4150 & 0.4828 &
-0.0700 & 0.4150 & -0.4828 & 0.0700 \\ \hline
-0.4150 & 0.4828 & -0.0700 & -0.4150 & 0.4828 & -0.0700 & 0.4150 & -0.4828 &
0.0700 & 0.4150 & -0.4828 & 0.0700 \\ \hline
-0.4150 & 0.4828 & -0.0700 & 0.4150 & -0.4828 & 0.0700 & 0.4150 & -0.4828 &
0.0700 & -0.4150 & 0.4828 & -0.0700 \\ \hline
0.2789 & 0.1300 & 0.4951 & 0.2789 & 0.1300 & 0.4951 & 0.2789 & 0.1300 &
0.4951 & 0.2789 & 0.1300 & 0.4951 \\ \hline
0.2789 & 0.1300 & 0.4951 & -0.2789 & -0.1300 & -0.4951 & 0.2789 & 0.1300 &
0.4951 & -0.2789 & -0.1300 & -0.4951 \\ \hline
0.2789 & 0.1300 & 0.4951 & 0.2789 & 0.1300 & 0.4951 & -0.2789 & -0.1300 &
-0.4951 & -0.2789 & -0.1300 & -0.4951 \\ \hline
0.2789 & 0.1300 & 0.4951 & -0.2789 & -0.1300 & -0.4951 & -0.2789 & -0.1300 &
-0.4951 & 0.2789 & 0.1300 & 0.4951 \\ \hline
\end{tabular}%
}
\end{table*}

\begin{table*}[h]
\caption{Example of spreading sequence of LCMA, $K$=16, $N_{S}$=8.}
\label{Table_SpreadingSequenceK12}\centering{\tiny
\begin{tabular}{|l|l|l|l|l|l|l|l|l|l|l|l|l|l|l|l|}
\hline
0 & 0.2789 & 0.4850 & 0.4702 & 0 & 0.2789 & 0.4850 & 0.4702 & 0 & 0.2789 &
0.4850 & 0.4702 & 0 & 0.2789 & 0.4850 & 0.4702 \\ \hline
0 & 0.2789 & 0.4850 & 0.4702 & 0 & -0.2789 & -0.4850 & -0.4702 & 0 & 0.2789
& 0.4850 & 0.4702 & 0 & -0.2789 & -0.4850 & -0.4702 \\ \hline
0 & 0.2789 & 0.4850 & 0.4702 & 0 & 0.2789 & 0.4850 & 0.4702 & 0 & -0.2789 &
-0.4850 & -0.4702 & 0 & -0.2789 & -0.4850 & -0.4702 \\ \hline
0 & 0.2789 & 0.4850 & 0.4702 & 0 & -0.2789 & -0.4850 & -0.4702 & 0 & -0.2789
& -0.4850 & -0.4702 & 0 & 0.2789 & 0.4850 & 0.4702 \\ \hline
0.5000 & 0.4150 & 0.1216 & -0.1700 & 0.5000 & 0.4150 & 0.1216 & -0.1700 &
0.5000 & 0.4150 & 0.1216 & -0.1700 & 0.5000 & 0.4150 & 0.1216 & -0.1700 \\
\hline
0.5000 & 0.4150 & 0.1216 & -0.1700 & -0.5000 & -0.4150 & -0.1216 & 0.1700 &
0.5000 & 0.4150 & 0.1216 & -0.1700 & -0.5000 & -0.4150 & -0.1216 & 0.1700 \\
\hline
0.5000 & 0.4150 & 0.1216 & -0.1700 & 0.5000 & 0.4150 & 0.1216 & -0.1700 &
-0.5000 & -0.4150 & -0.1216 & 0.1700 & -0.5000 & -0.4150 & -0.1216 & 0.1700
\\ \hline
0.5000 & 0.4150 & 0.1216 & -0.1700 & -0.5000 & -0.4150 & -0.1216 & 0.1700 &
-0.5000 & -0.4150 & -0.1216 & 0.1700 & 0.5000 & 0.4150 & 0.1216 & -0.1700 \\
\hline
\end{tabular}%
}
\end{table*}

Fig. \ref{Fig_FER_LCMA_LDPC_AWGNMA} shows the BLER of LCMA with $N_{S}=8$
and $K=16,20,24$, where a rate 1/2 length-3840 LDPC code in the 5G NR
standard and BPSK are used. The spreading matrices used for system loads
150\% and 200\% are shown in Table I and II, respectively\footnote{%
The spreading matrices for system loads 250\% and 300\% are too large to be
presented here. Interested readers may contact us to request the spreading
matrices for any system loads.}. We also compare to existing non-lattice
based IDMA system with chip-level interleaving and iterative elementary
signal estimation (ESE) detection, and SCMA system with iterative message
passing detection and decoding\footnote{%
We do not include comparison to RSMA, as it is for the close-loop MA system
where rate allocation is employed \cite{MaoTcom19}. We do not include PDMA,
MUSA and etc. for comparison, as their mechanisms are not largely different
to IDMA and SCMA.}. For a fair comparison, all MA schemes have identical
power among $K$ users.
\begin{figure}[h]
\centering
\includegraphics[width=0.42\textwidth]{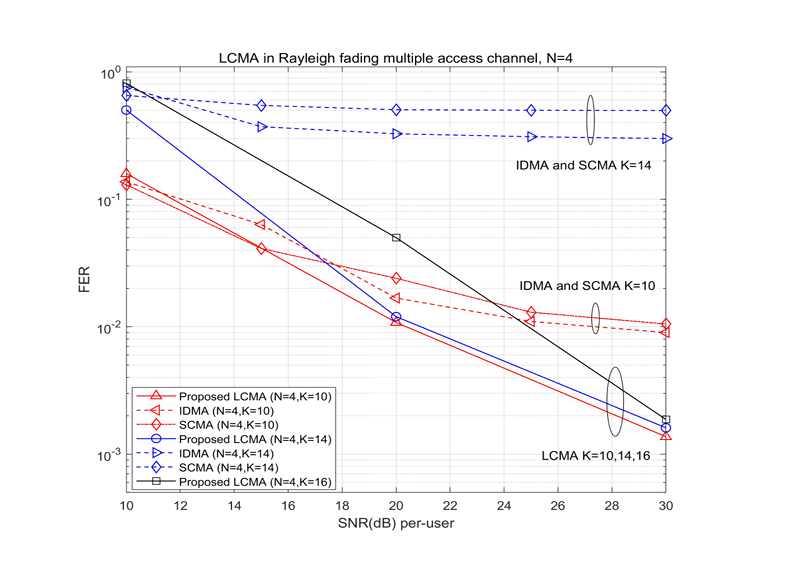}
\caption{BLER of LCMA with QPSK in fading MAC, $N=4$. }
\label{Fig_fadingMAC_LCMA}
\end{figure}
It is observed that LCMA exhibits the following advantages. First, LCMA
supports all system loads under consideration. In contrast, IDMA and SCMA
fail to support a system load greater than 200\%. This is due to the poor
adaptation of the 5G LDPC code with the ESE or message passing detector,
i.e., a poor convergence behavior of the iterative detection and decoding
(IDD) following the principle of EXIT chart. Due to the nature of parallel
processing of LCMA, the convergence problem is not relevant, thus the
stronger the underlying channel code is, the better the performance. We note
that such competitive performance is achieved with merely parallel
processing and $K$ single-user decoding, without using successive
cancelation or IDD. Second, excellent BLER performance of as low as $10^{-6}$
to $10^{-7}$ was demonstrated for LCMA, for all system loads under
consideration. In practice, the very low BLER may help with ultra-reliable
low-latency communication on top of massive access, by significantly
reducing requests for MU-ARQ retransmission. Lastly, it is shown that, the
supported system load by LCMA increase by 50\% for every SNR increase of
about 1.5 dB. Such consistent behavior was not reported for other existing
MA schemes in the literature, to the best of our knowledge.

 {Fig. \ref{Fig_fadingMAC_LCMA} shows the frame error performance in a fading MAC, where the spreading sequence developed for the Gaussian MAC is used. At the receiver side, the effective channel matrix is given by the combination of the spreading matrix $\mathbf{S}$ and the complex-valued channel gain $h_i,i=1,\cdots,K$. It is demonstrated that LCMA can support up to $K$=16 users with a spreading sequence length $N=4$, which considerably outperforms existing IDMA and SCMA schemes.}

\begin{figure}[h]
\centering
\includegraphics[width=0.4\textwidth]{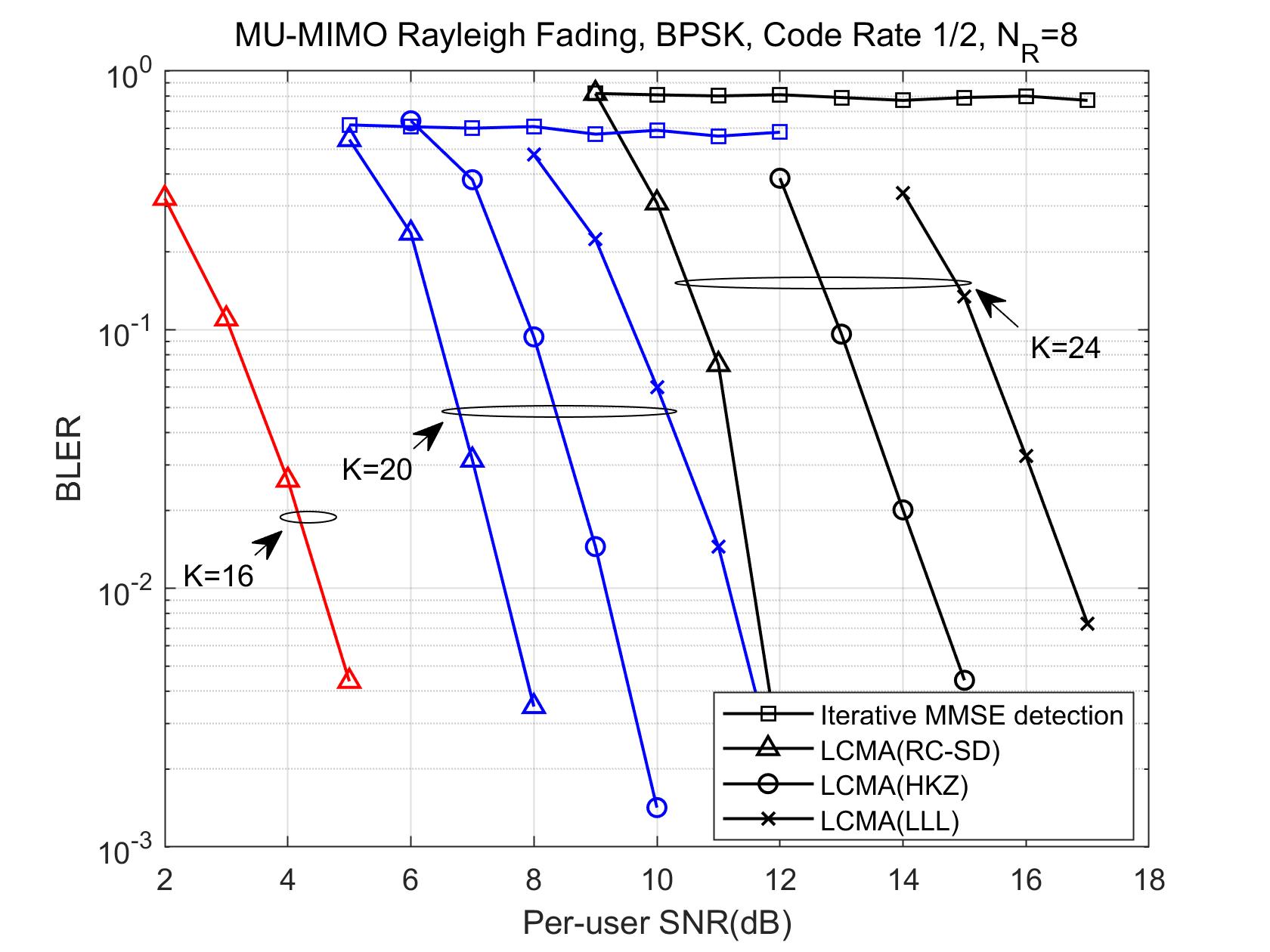}
\caption{BLER of LCMA in multi-user MIMO of $N_{R}=$8 receive antennas. LCMA
can support a system load of no less than 300\%, while the baseline scheme
with iterative receiver cannot support a load greater than 200\%.}
\label{Fig_FER_LCMA_MUMIMO_N8}
\end{figure}

\begin{figure}[h]
\centering
\includegraphics[width=0.4\textwidth]{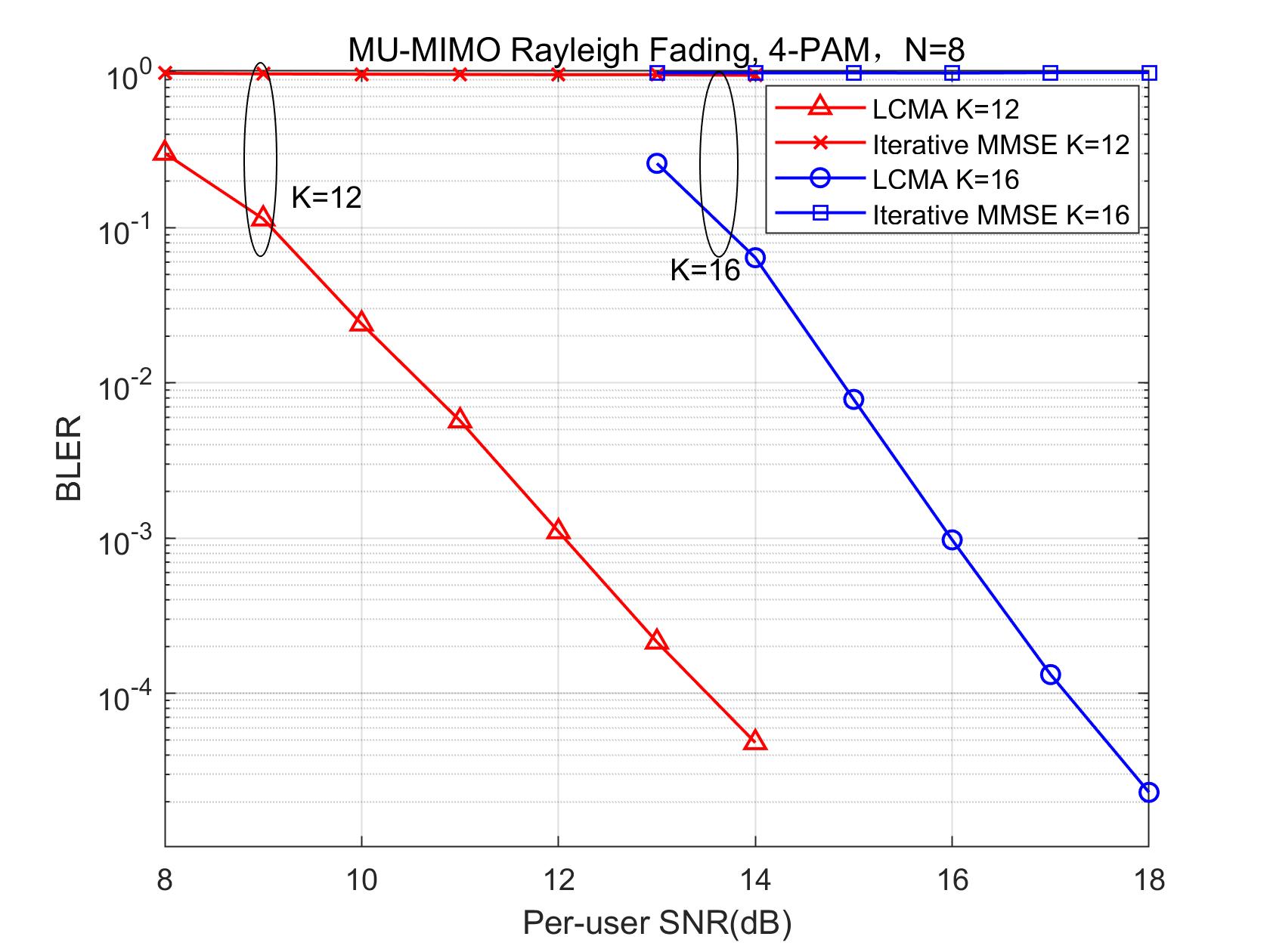}
\caption{BLER of LCMA with 4-PAM in MU-MIMO, $N_{R}=8$, $K$=12 and 16. For
4-PAM, LCMA can support a system load of at least 200\%, while iterative
MMSE detection fails to converge.}
\label{Fig_FER_LCMA_MUMIMO_N4_4PAM_LSD_withIDMASCMA}
\end{figure}

\subsection{LCMA for MU-MIMO}

Here we present the numerical results for MU-MIMO where the
receiver is equipped with $N_{R} $ antennas. The channel coefficients follow
the Rayleigh distribution. In this setting, the iterative ESE or BP
algorithms are implemented in the form of an iterative linear MMSE soft
cancelation algorithm: the signal of each received antenna can be viewed as
a chip-level signal in IDMA/SCMA; the chip-level cancelation with elementary
extrinsic information feedback is conducted;the linear MMSE filtering
combines all $N_{R}$ received antennas' signals.

Fig. \ref{Fig_FER_LCMA_MUMIMO_N8} shows the BLER of LCMA where $N_{R}=8$ and
$K=16,20,24$. BPSK and a length-480 5G NR LDPC code of rate $k/m$=1/2 are
utilized. $Q=10$ receiver iterations are implemented in iterative MMSE
detection. Perfect receiver-side CSI is considered in the simulation. It is clear that LCMA can support a system load of no less than
300\%, while the baseline scheme with iterative receiver cannot support a
system load greater than 200\% where the BLER curve flats out. 

We next consider MA with higher level modulations, e.g., 2$^{m}$-PAM (or 2$%
^{2m}$-QAM). Each user utilizes the 2$^{m}$-ary ring code for encoding (\ref%
{Eq_encodinggeneral}) and mapped to 2$^{m}$-PAM. Fig. \ref%
{Fig_FER_LCMA_MUMIMO_N4_4PAM_LSD_withIDMASCMA} shows the BLER of LCMA with $%
m=2$ (4-PAM), code rate $k/n=1/2$ $N_{R}=8$, $K$=12 and 16. The information
rate is 12 and 16 bits per channel-use per real dimension, respectively. The
detail of the underlying ring code can be found in \cite{YuTcom22}. It is
demonstrated that for 4-PAM, LCMA can support a system load of at least
200\% with using only parallel processing. In contrast, iterative MMSE
detection with 4-PAM fails to converge at this system load. We again note
that a 2$^{m}$-ary ring-code is required in LCMA for 4-PAM. Existing schemes
such BICM and TCM are not lattice codes hence does not support the LCMA
processing.

\begin{figure}[h]
\centering
\includegraphics[width=0.4\textwidth]{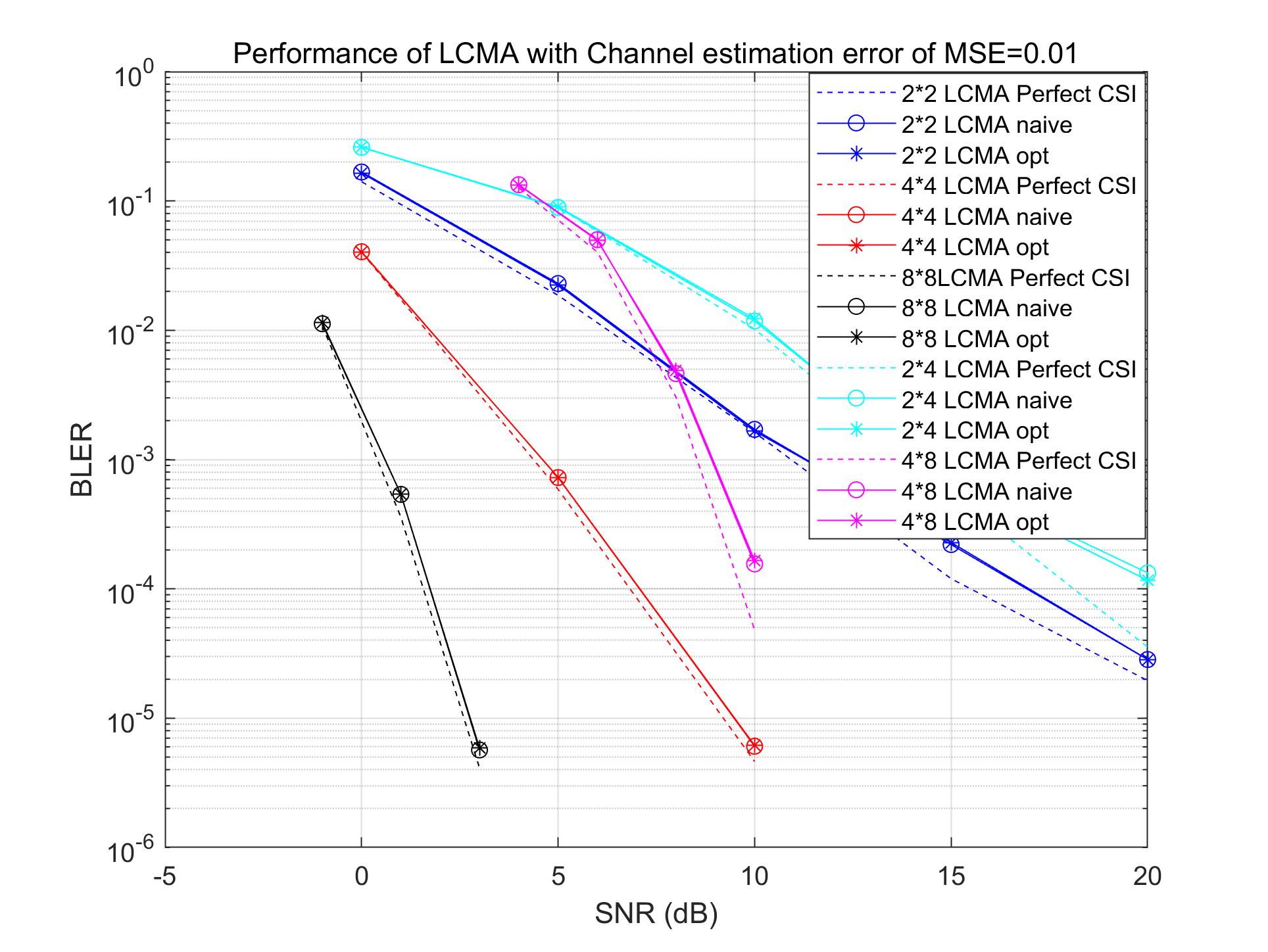}
\caption{BLER of LCMA with QPSK in MU-MIMO with imperfect CSI. }
\label{Fig_MUMIMO_LCMA_imperfectCSI}
\end{figure}

 {We note that any existing method for training sequence design and channel estimation algorithm for $\mathbf{H}$ can be adopted, such as the minimum mean square error (MMSE) and approximate message passing (AMP) methods. No extra requirement of channel estimation is needed for implement LCMA. Fig. \ref{Fig_MUMIMO_LCMA_imperfectCSI} shows the numerical results of LCMA with imperfect channel estimation where the MSE of channel estimation is 0.01. It is observed that as the dimension of the MA system increases, the loss due to channel estimation error becomes insignificant.}

\begin{figure}[h]
\centering
\includegraphics[width=0.4\textwidth]{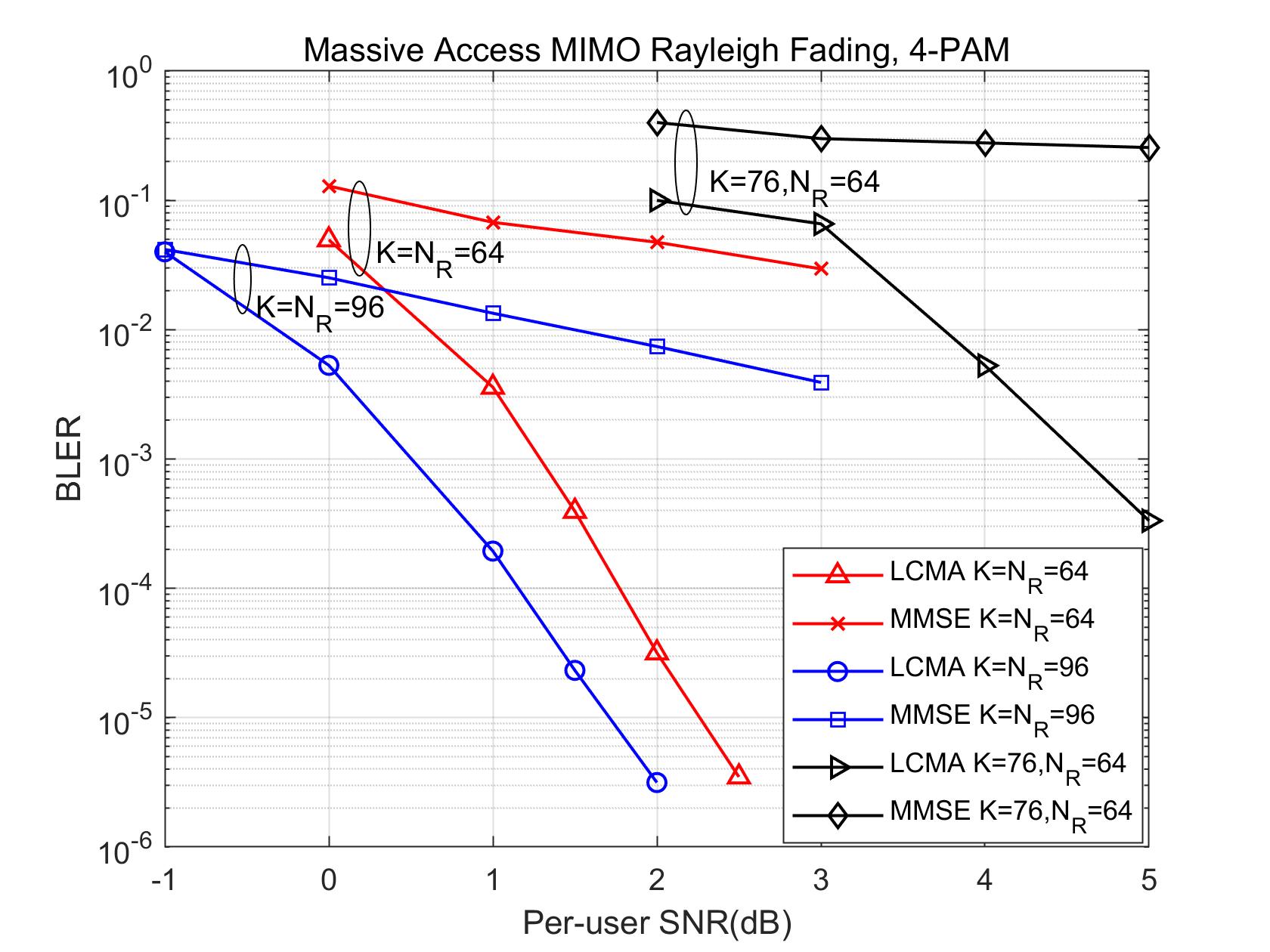}
\caption{BLER of LCMA in massive-access MIMO of $N_{R}=$64 and 96 and
receive antennas, where 4-PAM and the rate half 4-ary ring code is used.
LCMA vastly outperforms conventional MMSE receiver.}
\label{Fig_BLER_LCMA_MassiveMIMO_N64}
\end{figure}

\subsection{LCMA for Massive-Access MIMO}

Fig. \ref{Fig_BLER_LCMA_MassiveMIMO_N64} presents the numerical results for
the massive-access MIMO setup where the receiver is equipped with a large
number of antennas, i.e. $N_{R}$=64 and 96, while the number of UEs $K$ is
no smaller than $N_R$. In this setting, we compare to linear MMSE receiver.
It is observed that LCMA vastly outperforms the conventional baseline scheme
for the massive-access MIMO setup. LCMA can achieve a BLER as low as $%
10^{-6} $, which is for beyond the capability of the conventional scheme.
For the setup with $K=76,N_R=64$, the BLER of the baseline scheme flats out
while LCMA reaches $3\times10^{-4}$ at 5 dB.

\subsection{Analysis of Implementation Costs of LCMA}

\begin{table*}[h]
\caption{The orders of complexities of LCMA, IDMA and SCMA systems}
\label{table_complexity}%
\center{\tiny
\begin{tabular}{|c|c|c|c|c|}
\hline
& Detection & Decoding & Coefficient Identification & Interleaver\&De-interleaver \\ \hline
LCMA & $O\left( Knq \cdot E\left( \omega _{H}\left( \mathbf{a}\right)
\right) \right) $ for Det. Method II and III, $O(Kn |\mathcal{L}|)$ for Method I & $O(Kn\left( q-1\right) )$ & Between $O(K^{3}) $ and $O(K^{4})$ & not required \\ \hline
IDMA & $O(Q\cdot Knlog_{2}^{q}\cdot N_{S})$ & $O(Q\cdot Kn
log_{2}^{q})$ & not required & $O(2Q\cdot Kn\cdot N_{S})$
\\ \hline
SCMA & $O(Q\cdot K^{2}nlog_{2}^{q})$ & $O(Q\cdot Kn
log_{2}^{q})$ & not required & $O(2Q\cdot Kn )$ \\ \hline
\end{tabular}
}
\end{table*}

The orders of complexities are shown in Table. \ref{table_complexity}. The
typical value of receiver iterations $Q$ is between 4 to 10 for IDMA/SCMA.
The computation in LCMA consists of 1) channel-code decoding, 2) LCMA soft
detection, and 3) identification of $\mathbf{A}$. For 1), LCMA requires only
$K$ decoding operations while IDMA/SCMA requires $Q$ times more. For the
uplink MA, the modulation order $q=2^{m}$ is usually not large, where the
complexity of ring-code decoding is not considerably greater than that based
on binary channel code decoding.

For 2), LCMA needs to compute $K$ streams of APPs w.r.t. the ILCs, while
IDMA/SCMA requires to compute $Q\cdot K$ streams of soft estimates. In
particular, if Detection methods II or III is utilized, the per-symbol
detection complexity (of calculating the distance and the likelihood
function) of stream $l$ is of order $O(\left( q-1\right) \omega _{H}\left(
\mathbf{a}_{l}\right) )$, where $\omega _{H}\left( \mathbf{a}_{l}\right) <K$
denotes the weight of the coefficient vector $\mathbf{a}_{l}$. The average
detection complexity of LCMA is thus $O\left( Kn\left( q-1\right) E\left(
\omega _{H}\left( \mathbf{a}\right) \right) \right) $. In contrast, the
iterative chip-by-chip detection of IDMA has a complexity of $O\left( Q\cdot
Kn\log _{2}^{q}N_{S}\right) $, while that of SCMA is $O\left( Q\cdot Kn\log
_{2}^{q}E\left( \omega \left( \mathbf{s}\right) \right) \right) $ where $%
E\left( \omega \left( \mathbf{s}\right) \right) $ denotes the average number
of non-zero entries of the spreading sequence $\mathbf{s}$. Due to the
avoidance of iterative detection, the overall detection complexity of LCMA
is smaller than that of IDMA/SCMA.

For 3), with LLL, the complexity is between $O(K^{3})$ and $O(K^{4})$, a
polynomial in $K$. The complexity of HKZ and RC-SD is moderately higher than
LLL. Since $\mathbf{A}$ is chosen once per block, for a moderate-to-long
block length $n$ (e.g. $n>480$), this overhead is not significant.
\begin{figure}[h]
\centering
\includegraphics[width=0.35\textwidth]{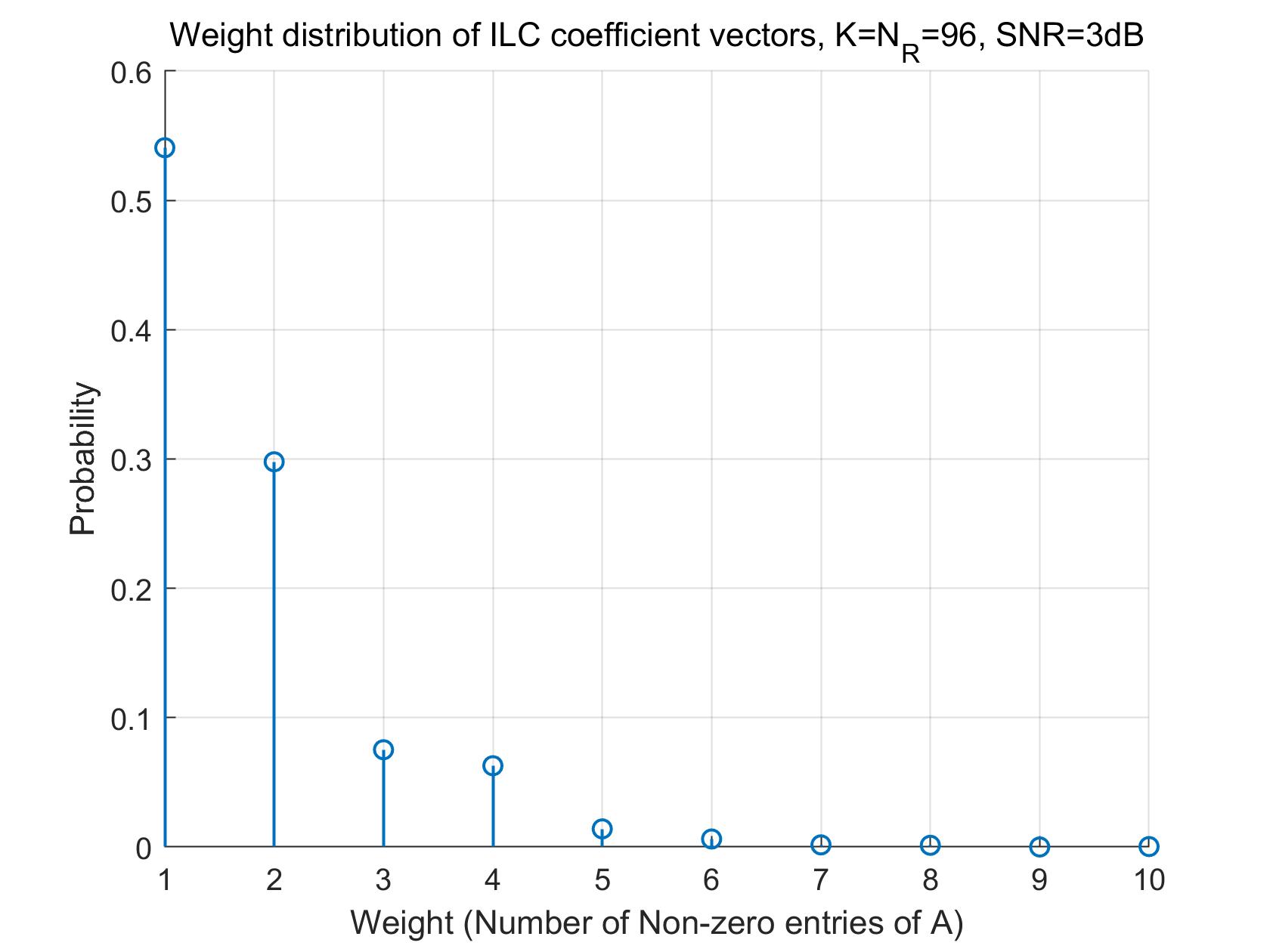}
\caption{Weight distribution of the ILC coefficient vector for
massive-access MIMO of $K=N_{R}=96$. }
\label{Fig_WeightDistribution_LCMA_MassiveMIMO_N96}
\end{figure}
We next discuss the complexity of LCMA for massive-access MIMO. Fig. \ref%
{Fig_WeightDistribution_LCMA_MassiveMIMO_N96} shows the distribution of the
number of non-zeros entries (i.e. the weight) of the ILC coefficient vectors
$\mathbf{a}$. This figure means that, although the number of UEs $K$ is
quite large, $\mathbf{A}$ is a reasonably sparse matrix, i.e, more than 90\%
ILC coefficient vectors are of weights less than 5. Recall from Section IV
that the average per-user complexity has order $O\left( 2^{m}E_{\mathbf{a}%
}\left( \omega _{H}\left( \mathbf{a}\right) \right) \right) $. This is $%
E\left( \omega _{H}\left( \mathbf{a}\right) \right) $ times of the
complexity of single-user detection. Fig. \ref%
{Fig_WeightDistribution_LCMA_MassiveMIMO_N96} shows that $E_{\mathbf{a}%
}\left( \omega _{H}\left( \mathbf{a}\right) \right) $ is just a very small
fraction of $K$ in general. As such, LCMA has a very low complexity for
massive access MIMO.

There are other features that may be desired for implementation. First, note
that the $Q$ consecutive receiver iterations in IDMA/SCMA are executed in
serial. Since IDD is avoided, LCMA with parallel processing incurs a much
smaller latency. Second, there is no need to store the chip-level soft
extrinsic information updated in IDD, which may reduce the memory
occupation. Last but not that least, for BPSK/QPSK, off-the-shelf channel
codes in various standards can be directly used in LCMA, regardless of the
system load $\frac{K}{N}$. In contrast, as the load $\frac{K}{N}$ varies,
IDMA and SCMA have to adopt different codes. Otherwise, the convergence of
IDD may not be achieved, leading to impaired performance or even failed
functionality.

\section{Conclusions}

We presented a LCMA scheme and a package of algorithms that are
essential to its implementation, including the $2^{m}$-ary ring-coded
PAM, LCMA soft detection algorithms, rate-constrained sphere-decoding for
solving the BIVP that identifies the optimized coefficient matrix $\mathbf{A}
$, and a pragmatic solution for optimizing the MA spreading matrix $\mathbf{S%
}$. The per-user detection complexity is of order less than $O(K)$, suitable
for massive access. With just parallel processing, considerable system load
and error rate performance enhancement were demonstrated, without using successive interference cancelation or iterative
detection. Off-the-shelf binary codes such as 5G NR LDPC codes can be
directly used in LCMA for any system load, avoiding the issue of adaptation
of channel-code and multi-user detector. Excellent BLER performance of as
low as $10^{-6}$ to $10^{-7}$ was demonstrated for LCMA, for all system
loads under consideration. Such very low BLER may help with ultra-reliable
low-latency communication on top of massive access, by tremendously slashing
request of MU-ARQ retransmission.


\bibliographystyle{ieeetr}
\bibliography{RefTaoYang}

\begin{thebibliography}{10}

\bibitem{6GTextbookTongWen}
W.~Tong and P.~Zhu, ``{6G} the next horizon-from connected people and things to
  connected intelligence,'' {\em Cambridge University Press}, 2021.

\bibitem{Dai18}
L.~Dai, B.~Wang, Z.~Ding, Z.~Wang, S.~Chen, and L.~Hanzo, ``A survey of
  non-orthogonal multiple access for {5G},'' {\em IEEE Communications Surveys
  and Tutorials}, vol.~20, no.~3, pp.~2294--2323, 2018.

\bibitem{ding2017application}
Z.~Ding, Y.~Liu, J.~Choi, Q.~Sun, M.~Elkashlan, I.~Chih-Lin, and H.~V. Poor,
  ``Application of non-orthogonal multiple access in {LTE} and {5G} networks,''
  {\em IEEE Commun. Mag.}, vol.~55, no.~2, pp.~185--191, 2017.

\bibitem{CoverTextbook}
T.~M. Cover and J.~A. Thomas, ``Elements of information theory,'' {\em John
  Wiley \& Sons, Inc.}, 1991.

\bibitem{ten2003design}
S.~ten Brink and G.~Kramer, ``Design of repeat-accumulate codes for iterative
  detection and decoding,'' {\em IEEE Trans. Signal Processing}, vol.~51,
  no.~11, pp.~2764--2772, 2003.

\bibitem{islam2016power}
S.~R. Islam, N.~Avazov, O.~A. Dobre, and K.-S. Kwak, ``Power-domain
  non-orthogonal multiple access {(NOMA)} in {5G} systems: {Potentials and
  challenges},'' {\em IEEE Commun. Surv. Tuts.}, vol.~19, no.~2, pp.~721--742,
  2016.

\bibitem{wang1999iterative}
X.~Wang and H.~V. Poor, ``Iterative (turbo) soft interference cancellation and
  decoding for coded {CDMA},'' {\em IEEE Trans. Comm.}, vol.~47, no.~7,
  pp.~1046--1061, 1999.

\bibitem{LiTWC06}
P.~Li, L.~Liu, K.~Wu, and W.~Leung, ``Interleave division multiple-access,''
  {\em IEEE Trans. Wireless Comm.}, vol.~5, no.~4, pp.~938--947, Apr. 2006.

\bibitem{nikopour2013sparse}
H.~Nikopour and H.~Baligh, ``Sparse code multiple access,'' in {\em Proc. IEEE
  24th Int. Symp. Pers. Indoor Mobile Radio Commun. (PIMRC)}, pp.~332--336,
  2013.

\bibitem{KudekarISIT11}
S.~Kudekar and K.~Kasai, ``Spatially coupled codes over the multiple access
  channel,'' in {\em 2011 IEEE International Symposium on Information Theory
  Proceedings}, pp.~2816--2820, 2011.

\bibitem{YangTWC09}
T.~Yang, J.~Yuan, and Z.~Shi, ``Rate optimization for {IDMA} systems with
  iterative joint multi-user decoding,'' {\em IEEE Trans. Wireless Comm.},
  vol.~8, no.~3, pp.~1148--1153, Mar. 2009.

\bibitem{RimoldiIT96}
B.~Rimoldi and R.~Urbanke, ``A rate-splitting approach to the {Gaussian}
  multiple-access channel,'' {\em IEEE Transactions on Inf. Theory}, vol.~42,
  no.~2, pp.~364--375, 1996.

\bibitem{MaoTcom19}
Y.~Mao, B.~Clerckx, and V.~O.~K. Li, ``Rate-splitting for multi-antenna
  non-orthogonal unicast and multicast transmission: Spectral and energy
  efficiency analysis,'' {\em IEEE Trans. Comm.}, vol.~67, no.~12,
  pp.~8754--8770, 2019.

\bibitem{chen2016pattern}
S.~Chen, B.~Ren, Q.~Gao, S.~Kang, S.~Sun, and K.~Niu, ``Pattern division
  multiple access--{A} novel nonorthogonal multiple access for fifth-generation
  radio networks,'' {\em IEEE Trans. Vehi. Tech.}, vol.~66, no.~4,
  pp.~3185--3196, 2016.

\bibitem{ChengJSAC21}
Y.~Cheng, L.~Liu, and L.~Ping, ``Orthogonal {AMP} for massive access in
  channels with spatial and temporal correlations,'' {\em IEEE J. Sel. Areas
  Commun.}, vol.~39, no.~3, pp.~726--740, 2021.

\bibitem{SunTcom17}
Z.~Sun, Y.~Xie, J.~Yuan, and T.~Yang, ``Coded slotted aloha for erasure
  channels: Design and throughput analysis,'' {\em IEEE Trans. Commun.},
  vol.~65, no.~11, pp.~4817--4830, 2017.

\bibitem{PaoliniIT15}
E.~Paolini, G.~Liva, and M.~Chiani, ``Coded slotted aloha: A graph-based method
  for uncoordinated multiple access,'' {\em IEEE Trans. Inf. Theory}, vol.~61,
  no.~12, pp.~6815--6832, 2015.

\bibitem{ZamirTIT02}
R.~{Zamir}, S.~{Shamai}, and U.~{Erez}, ``Nested linear/lattice codes for
  structured multiterminal binning,'' {\em IEEE Trans. Inf. Theory}, vol.~48,
  no.~6, pp.~1250--1276, 2002.

\bibitem{NazerIT11}
B.~Nazer and M.~Gastpar, ``Compute-and-forward: Harnessing interference through
  structured codes,'' {\em IEEE Trans. Inf. Theory}, vol.~57, no.~10,
  pp.~6463--6486, Oct. 2011.

\bibitem{NamIT10}
W.~Nam, S.~Chung, and Y.~H. Lee, ``Capacity of the {G}aussian two-way relay
  channel to within 1/2 bit,'' {\em IEEE Trans. Inf. Theory}, vol.~56, no.~11,
  pp.~5488--5494, Nov. 2010.

\bibitem{YuanYangIT13}
X.~Yuan, T.~Yang, and I.~Collings, ``Multiple-input multiple-output two-way
  relaying: a space-division approach,'' {\em IEEE Trans. Inf. Theory},
  vol.~59, no.~10, pp.~6421--6440, Oct. 2013.

\bibitem{LimTIT20}
S.~H. Lim, C.~Feng, A.~Pastore, B.~Nazer, and M.~Gastpar, ``Compute-forward for
  {DMCs}: Simultaneous decoding of multiple combinations,'' {\em IEEE Trans.
  Inf. Theory}, vol.~66, no.~10, pp.~6242--6255, 2020.

\bibitem{ZhanIT14}
J.~Zhan, B.~Nazer, U.~Erez, and M.~Gastpar, ``Integer-forcing linear
  receivers,'' {\em IEEE Trans. Inf. Theory}, vol.~60, no.~12, pp.~7661--7685,
  Dec. 2014.

\bibitem{SilvaTWC17}
D.~Silva, G.~Pivaro, G.~Fraidenraich, and B.~Aazhang, ``On integer-forcing
  precoding for the {Gaussian} {MIMO} broadcast channel,'' {\em IEEE Tran.
  Wireless Comm.}, vol.~16, no.~7, pp.~4476--4488, 2017.

\bibitem{HongIT13}
S.-N. Hong and G.~Caire, ``Compute-and-forward strategies for cooperative
  distributed antenna systems,'' {\em IEEE Trans. Inf. Theory}, vol.~59, no.~9,
  pp.~5227--5243, Sep. 2013.

\bibitem{YangTWC17}
T.~Yang, ``Distributed {MIMO} broadcasting: Reverse compute-and-forward and
  signal-space alignment,'' {\em IEEE Trans. Wireless Comm.}, vol.~16, no.~1,
  pp.~581--593, 2017.

\bibitem{SakzadTWC13}
A.~Sakzad, J.~Harshan, and E.~Viterbo, ``Integer-forcing {MIMO} linear
  receivers based on lattice reduction,'' {\em IEEE Trans. Wireless Comm.},
  vol.~12, no.~10, pp.~4905--4915, 2013.

\bibitem{YangTWC20}
D.~Yang and K.~Yang, ``Multimode integer-forcing receivers for block fading
  channels,'' {\em IEEE Transactions on Wireless Communications}, vol.~19,
  no.~12, pp.~8261--8271, 2020.

\bibitem{LyuTIT19}
S.~Lyu, A.~Campello, and C.~Ling, ``Ring compute-and-forward over block-fading
  channels,'' {\em IEEE Transactions on Information Theory}, vol.~65, no.~11,
  pp.~6931--6949, 2019.

\bibitem{OrdentlichTIT12}
O.~Ordentlich and U.~Erez, ``Cyclic-coded integer-forcing equalization,'' {\em
  IEEE Transactions on Information Theory}, vol.~58, no.~9, pp.~5804--5815,
  2012.

\bibitem{zhu2016gaussian}
J.~Zhu and M.~Gastpar, ``Gaussian multiple access via compute-and-forward,''
  {\em IEEE Trans. Inf. Theory}, vol.~63, no.~5, pp.~2678--2695, 2016.

\bibitem{sula2018compute}
E.~Sula, J.~Zhu, A.~Pastore, S.~H. Lim, and M.~Gastpar, ``Compute--forward
  multiple access {(CFMA)}: {Practical} implementations,'' {\em IEEE Trans.
  Comm.}, vol.~67, no.~2, pp.~1133--1147, 2018.

\bibitem{ChenISIT22}
Q.~Chen, F.~Yu, T.~Yang, J.~Zhu, and R.~Liu, ``A linear physical-layer network
  coding based multiple access approach,'' in {\em 2022 IEEE International
  Symposium on Information Theory (ISIT)}, pp.~2803--2808, 2022.

\bibitem{ChenTWC22}
Q.~Chen, F.~Yu, T.~Yang, and R.~Liu, ``Gaussian and fading multiple access
  using linear physical-layer network coding,'' {\em IEEE Trans. Wireless
  Comm.}, early access,2022.

\bibitem{YangTWC17_NoMA}
T.~Yang, L.~Yang, Y.~J. Guo, and J.~Yuan, ``A non-orthogonal multiple-access
  scheme using reliable physical-layer network coding and cascade-computation
  decoding,'' {\em IEEE Trans. Wireless Comm.}, vol.~16, no.~3, pp.~1633--1645,
  2017.

\bibitem{ErezTIT05}
U.~{Erez} and S.~{ten Brink}, ``A close-to-capacity dirty paper coding
  scheme,'' {\em IEEE Trans. Inf. Theory}, vol.~51, no.~10, pp.~3417--3432,
  2005.

\bibitem{YuTcom22}
F.~Yu, T.~Yang, and Q.~Chen, ``Doubly irregular repeat modulation codes over
  integer rings for multi-user communications,'' {\em ChinaComm.},
  2022,accepted.

\bibitem{ping2006interleave}
L.~Ping, L.~Liu, K.~Wu, and W.~K. Leung, ``Interleave division
  multiple-access,'' {\em IEEE Trans. Wireless Commun.}, vol.~5, no.~4,
  pp.~938--947, 2006.

\bibitem{YangTaoCL23}
T.~Yang, ``Beyond integer-forcing receiver for lattice-code based multi-user
  {MIMO} system,'' {\em IEEE Communications Letters}, vol.~27, no.~10,
  pp.~2553--2557, 2023.

\bibitem{LenstraLLL}
H.~W. J. L.~L. Lenstra, A. K.;~Lenstra, ``Factoring polynomials with rational
  coefficients,'' {\em Mathematische Annalen}, vol.~261, no.~4, pp.~515--534,
  1982.

\bibitem{LyuTSP17}
S.~Lyu and C.~Ling, ``Boosted {KZ} and {LLL} algorithms,'' {\em IEEE
  Transactions on Signal Processing}, vol.~65, no.~18, pp.~4784--4796, 2017.

\bibitem{ErezTIT04}
U.~{Erez} and R.~{Zamir}, ``Achieving log(1+{SNR}) on the {AWGN} channel with
  lattice encoding and decoding,'' {\em IEEE Trans. Inf. Theory}, vol.~50,
  no.~10, pp.~2293--2314, 2004.

\bibitem{FangGlobecom16}
D.~Fang, Y.-C. Huang, Z.~Ding, G.~Geraci, S.-L. Shieh, and H.~Claussen,
  ``Lattice partition multiple access: A new method of downlink non-orthogonal
  multiuser transmissions,'' in {\em 2016 IEEE Global Communications Conference
  (GLOBECOM)}, pp.~1--6, 2016.

\end{thebibliography}

\end{document}